\documentstyle[lprocl,11pt]{article}  
\def\Journal#1#2#3#4{{#1} {\bf #2}, #3 (#4)}
\def\Journalm#1#2#3#4{{{\em #1}} {\bf #2}, #3 (#4)}
\def\vec#1{{\bf #1}}  
\def\primeq{\def\theequation{\arabic{equation}'}}  
\def\normeq{\def\theequation{\arabic{equation}}}  
\def\matr#1{{\underline {\underline {{\bf #1}}}}}  
\def\PRL{\em Phys.\ Rev.\ Lett.}
\def\PRB{{\em Phys.\ Rev.}~B}
\def\PRE{{\em Phys.\ Rev.}~E}

\def\JPCSSP{{\em J.\ Phys.}~C: {\em Solid State Phys.}}
\def\JSP{\em J.\ Stat.\ Phys.}
\def\JPAMG{{\em J.\ Phys.}~ A: {\em Math.\ Gen.}}
\def\be{\begin{equation}}
\def\ee{\end{equation}}
\def\bea{\begin{eqnarray}}
\def\eea{\end{eqnarray}}
  
\begin{document}  
\title{BEYOND THE SHERRINGTON-KIRKPATRICK MODEL}
\author{ C. DE DOMINICIS }
\address{Service de Physique Th\'eorique, CEA Saclay,\\
F-91191 Gif sur Yvette, France}
\author{ I. KONDOR }
\address{Bolyai College, E\"otv\"os University,\\
H-1145 Amerikai \'ut 96, Budapest, Hungary}
\author{ T. TEMESV\'ARI }
\address{Institute for Theoretical Physics, E\"otv\"os University,\\
H-1088 Puskin utca 5-7, Budapest, Hungary}

\maketitle\abstracts{
The state of art in spin glass field theory is reviewed.
We start from an Edwards-Anderson-type model in finite dimensions, with 
finite but long range forces, construct the effective field theory that 
allows one to extract the long wavelength behaviour of the model, and set 
up an expansion scheme (the loop expansion) in the inverse range of the 
interaction.
At the zeroth order we recover mean field theory. We evaluate systematic 
corrections to this around Parisi's replica symmetry broken solution.
At the level of quadratic fluctuations we derive a set of coupled 
integral equations for the free propagators of the theory and show how 
they can be solved for short, intermediate and extreme long distances.
To reveal the physical meaning of these results, we relate the 
various propagator components to overlaps of spin-spin correlation functions 
inside a single phase space valley resp.~between different valleys.
Next we calculate the first loop corrections to the theory 
above 8 dimensions, where we find that it maps back onto mean 
field theory, with basically temperature independent renormalization 
of the coupling constants, thereby demonstrating that Parisi's mean field 
theory is, at least perturbatively, stable against finite range corrections.
In the range between six and eight dimensions various physical quantities 
pick up nontrivial temperature dependences which can, however, still be 
determined exactly. Upon approaching the upper critical dimension ($d=6$) of 
the model, scaling which is badly violated in Parisi's mean field theory 
is gradually restored. 
Below 6 dimensions one should apply renormalization group methods.
Unfortunately, the structure of RG is not completely understood in spin
glass theory. Nevertheless, the first corrections in $6-d$ to e.g.\ the
exponent of the order 
parameter can still be calculated, moreover exponentiation to this power 
can be checked at the next order. The theory is, however, plagued by 
infrared divergences due to the presence of zero modes and soft modes.
Systematic methods (like those developed in the ${\rm O}(n)$ model) to handle 
these infrared singularities are not yet available in spin glass
theory.} 

\section{Introduction}  
  
Parisi's mean field theory  (MFT) [1]
is generally accepted as  
the correct solution of the Sherrington-Kirkpatrick  (SK)  
problem [2]. The crucial technical assumption that renders the  
SK model soluble is that every spin interacts with infinitely  
many neighbours. Such a situation could arise e.g.~in a system  
in infinitely high  spatial dimensions $d\to \infty$ or in a  
finite  
dimensional system with an infinitely long ranged interaction.  
While in some extensions of the theory (especially in applications   
outside physics, like combinatorial optimization or neural  
networks) the  
assumption of high connectivity is well justified, in  
real physical systems that live in low dimensions and have  
finite range forces MFT can serve, at best, as a zeroth  
approximation.  
  
In order to go beyond the SK model and to approach a more  
realistic situation, one can set up an expansion scheme either  
in the inverse dimensionality  ($1/d$ expansion) or in the  
inverse interaction radius (loop expansion). These will work,  
however, only if the phase space structure of the finite  
dimensional, finite ranged system is similar to that of the  
fully connected mean field model.  
  
 The  physical picture underlying Parisi's solution  is 
of the system having a rugged free-energy surface, with many,  
hierarchically organized equilibrium states in the frozen  
phase. However, a general consensus on whether such a structure can  
survive in real, finite dimensional, finite ranged models has  
never been reached. To the variety  of approaches in this decade-long 
debate, ranging from scaling [3,4]  
and phenomenological renormalization [5] to large scale  
simulations [6-13] and to the first perturbative steps within the  
$1/d$ [14] and the loop expansions [15,16], respectively, there have  
recently been added the exact methods of mathematical  
physics [17-21]. With this, a fundamental problem, namely that of  
the definition of an equilibrium state in a random system, has  
been brought into focus.  
  
The results to be reviewed in this paper have been obtained  
under the assumption that the ``many valley structure" of MFT  
remains relevant for a finite dimensional,  
finite ranged system. We use the formalism of replica field  
theory [1], calculate free propagators and set out to determine  
the first loop corrections. All our results will be confined to  
the Ising spin glass just below its freezing  temperature.  
  
The technical difficulties we encounter are considerable, so we  
can fully compute the first corrections to the order parameter  
and to the excitation spectrum only in high  ($d>6$)  
dimensions.  
Near $d=6$, the upper critical dimension, we have only partial  
results. We also notice severe infrared (IR) problems here.  
These are manifestations of the extreme sensitivity of the  
system to changes in distant regions: the unusually strong IR  
singularities in our propagators may well tell, in the language  
of replica field theory, the same story as the mathematical  
results concerning the chaotic response to changes on the  
boundary [20]. In this context we find it remarkable that the  
worst IR powers are displayed by those propagator components  
that also describe the chaotic response to infinitesimal  
variations in the control parameters [22,23].  
  
If these IR problems turn out to be unsurmountable they will  
destroy replica symmetry breaking (RSB) field theory from  
within. If, on the other hand, recent efforts by Parisi and  
coworkers [24] and also by  ourselves [25] succeed in saving the  
theory through formally exact statements like Ward identities  
etc., we will find it hard to believe that the theory is  
completely devoid of physical meaning.  
  
We interpret the goal of this paper in a very  
restricted sense: apart from citing the results by  
Georges, M\'ezard and Yedidia [14] in a qualitative manner we  
shall not be able to cover the method of $1/d$ expansion, 
nor the important set of papers by Parisi and coworkers on the  
finite size corrections to MFT [26], although both of these  
approaches might find their well deserved place under the title  
of this paper. What we will try to give an account of is the  
construction of the loop expansion for spin glasses below the  
freezing temperature. Within the space available we will have  
to be fairly  sketchy on this, too. We will not provide proofs  
but will only give some hints as to how the results we display  
can be derived. Concerning the results themselves, especially those for  
the propagators, we will try to be rather more exhaustive,  
however, and collect the various propagators, their limiting  
forms, etc.\ that have appeared over the years scattered in a  
series of papers [27-30,24]. We will also try to demonstrate their use  
for the calculation of short range corrections in a few cases. 
  
The plan of the paper is the following. In Sec.~2 we briefly  
review the elements of the field theoretic formulation of the  
problem. In Sec.~3 we give a sketchy account of Parisi's MFT  
with the purpose of fixing notation. We take the first step  
beyond MFT in Sec.~4: we spell out the set of equations for the  
free propagators and, for the sake of a first  
orientation, solve them under the (wrong) assumption of replica symmetry,  
thereby displaying the famous de Almeida-Thouless (AT) instability [31],
and also the two characteristic  ``mass scales" inherent in  
the system. In Sec.~5  we abandon the assumption of replica  
symmetry but confine our attention to the ``high-momentum"  
region near the upper cutoff where the equations for the  
propagators can still be solved by a simple iteration. The  
results obtained in this region will be used later when  
calculating the first loop  corrections in high dimensions  
($d>8$).  
In  Sec.~6 we investigate the propagators in the near infrared  
region, i.e.\ for momenta around the ``large mass" (which  
vanishes like the square root of the reduced temperature near  
the spin glass transition) and much larger than the ``small  
mass" (vanishing like the first power of the reduced  
temperature). We show that in this region the complicated set  
of (ten) integral equations for the propagators reduces to a  
set of linear algebraic equations, which can then be solved by  
elementary means. We turn to the general study of the  
formidable set of integral equations  for the propagators in  
Sec.~7. Their solution is broken down into two  steps. First, by  
exploiting the residual (ultrametric) symmetry of the system,  
we indicate how the problem of the inversion of a general 
ultrametric matrix can be reduced to the inversion of a much  
simpler object that we call the kernel. Next, in the  special case  
of the Hess matrix we show that in the vicinity of the  
transition temperature it has a very simple kernel which can be  
inverted  with relative ease and yields the propagators in  
closed form. The singularities of the propagators determine the  
excitation spectrum of the system, and show that Parisi's  
solution is marginally stable. From the complicated exact  
expressions for the propagators we extract the limiting forms,  
valid in the far infrared or extreme long wavelength limit  
(where momenta are comparable to or smaller than the small mass  
scale) in Sec.~8. The propagator components are related to some  
combinations of correlation functions. This relationship is  
established in Sec.~9 where the physical meaning of some of  
the results obtained so far is also analysed. We step beyond the  
analysis of Gaussian fluctuations in Sec.~10 where we derive the  
first loop corrections to the order parameter and to the  
excitation spectrum above 8 dimensions. By absorbing the loop  
corrections into the coupling constants we map back the theory  
onto MFT and thereby show  that Parisi's solution is, at least  
in high dimensions, perturbatively stable. However, the renormalized  
quartic coupling blows up  as one approaches $d=8$  
from  
above. Therefore in the range $6<d<8$ one has to rearrange the  
loop  
expansion. This is done is Sec.~11. At and below $d=6$ the loop  
expansion is bound to break down completely. Although the  
structure of the renormalization group, the usual remedy in  
such a situation, is not known in  spin glass theory, we show in  
Sec.~12 that some information can still be extracted from the  
logarithmic singularities appearing here. In particular, we  
show that to first order in  $\varepsilon=6-d$ the critical  
exponent  $\beta$ can  
be computed in agreement with the values of other critical  
exponents known from above $T_c$ [32-34],  moreover, this power  
can be checked to exponentiate at the next order in  
$\varepsilon$. Sec.~13
concludes the paper with a brief summary.  
  
\section{The elements of replica field theory}  
  
Our starting point is a standard Edwards-Anderson-like [35]  
model  
for $N$ Ising spins in $d$  dimensions, with a long but  
finite-ranged interaction:  
\begin{eqnarray}  
{{\cal H}}=-\sum_{(i,j)}\frac{J_{ij}}{\sqrt z}\:f\left (\frac{  
| {{\bf r}}_i -{{\bf r}}_j |}{\rho a}\right )s_i  
s_j\quad .\end{eqnarray}  
  
In Eq.~1 the external field has been set to zero, the summation  
is over the pairs ($i$,$j$) of lattice sites, $z=\rho ^d$ is  
the  
number of spins within the interaction radius     measured in  
units  
of the lattice spacing $a$. $f(x)$ is a smooth positive  
function which takes the value 1 for $x\leq 1$,  
and decays to zero sufficiently fast for $x>1$, thereby  
cutting off the interaction around    $r \sim \rho a$.  
$J_{ij}$ are independent, Gaussian distributed random variables  
with  
mean  
zero and variance $\Delta ^2 $.  We will be interested in the  
long  
distance behaviour of the model (1), for which the details of  
$f(x)$ are to a large extent immaterial, so we  
can choose  $f$ according to convenience.  
  
The long wavelength properties near $T_c$ can be extracted by  
studying the associated effective Lagrangean of replica field  
theory [36] which, for an appropriate  
choice of $f$, works out to have the form:  
\begin{eqnarray}  
{{\cal L}}&=&-\frac{1}{4}\sum _{\alpha ,\beta }\sum _{  
\vec{p}} \left ((pa\rho )^2-2\tau\right ) |\Phi_{{\bf  
p}}^{\alpha  
\beta}|^2+\frac{w}{6\sqrt N}\sum _{ \alpha ,\beta ,\gamma }\sum  
 _{\vec {p_i}}\Phi ^{ \alpha \beta }_{\vec {p_1}}\Phi  
^{\beta \gamma }_{\vec {p_2}}\Phi ^{\gamma \alpha  
}_{\vec {p_3}}+ \nonumber \\  
&&+ \frac{u}{12N}\sum _{\alpha ,\beta }\sum  
_{\vec {p_i}}\Phi ^{ \alpha \beta }_{\vec  
{p_1}}\Phi ^{ \alpha \beta }_{\vec {p_2}}\Phi ^{ \alpha  
\beta }_{\vec {p_3}}\Phi ^{ \alpha \beta }_{\vec {p_4}}  
+\ldots ,  
\end{eqnarray}  
where the wavevector summations are restricted by ``momentum  
conservation" $\sum _i{\vec {p_i}}=0$ and by an UV cutoff 
at $p\sim 1/\rho a$.  
$\tau $ is the reduced temperature measured relative to the  
mean  
field value $T_c^{MF}=\Delta $ of the critical temperature:  
\begin{eqnarray}  
\tau =\frac{T_c^{MF}-T}{T_c^{MF}}\quad .  
\end{eqnarray}  
$\tau$  will be assumed small throughout this paper.  
  
The fields $\Phi ^{ \alpha \beta }$ are symmetric in the  
replica  
indices $\alpha ,\beta =1,2,\ldots ,n$ and $\Phi ^{\alpha  
\alpha  
}\equiv 0$.  
In the truncated Lagrangean (2), first proposed by Parisi  [1]  
and used by several authors since, we have kept only that  
particular quartic term which is responsible for replica  
symmetry breaking on the mean field level.  
The numerical  values of the bare coupling constants $w$ and  
$u$  
work out to be 1 in the case of the Ising spin glass. By a  
slight generalization of the model we wish to regard these bare  
couplings as  essentially free (positive) parameters, partly  
for book-keeping purposes, but also because they pick up short  
range corrections anyhow.  
  
Now we split the field into an equilibrium and a fluctuating  
part as  
$$\Phi ^{ \alpha \beta }_{\vec {p}} = \sqrt Nq_{ \alpha \beta  
}\delta _{\vec {p},0}^{Kr}+\Psi _{\vec {p}} ^{ \alpha \beta }$$  
with  
\begin{eqnarray}  
{\begin{array}{rcl}  
\displaystyle q_{ \alpha \beta } &=&q_{\beta \alpha },  \\  
\Psi ^{ \alpha \beta }&=&\Psi ^{\beta \alpha },\\  
\end{array}}  
 & &{\begin{array}{rcl}  
q_{\alpha \alpha }&=&0 \\  
 \Psi ^{\alpha \alpha }&=&0\quad .\\  
\end{array}}  
\end{eqnarray}

The Lagrangean then splits into four terms:  
\begin{eqnarray}  
{{\cal L}}={{\cal L}}^{(0)}+{{\cal L}}^{(1)}+{{\cal  
L}}^{(2)}+{{\cal  
L}}^{(3)}+{{\cal L}}^{(4)}  
\end{eqnarray}  
defined as follows:  
\begin{eqnarray}  
{{\cal L}}^{(0)}&=&N\left [\frac{\tau }{2}\sum _{ \alpha, \beta  
}q^2_{ \alpha \beta }+\frac{w}{6}\sum _{ \alpha ,\beta , \gamma  
}q^{ \alpha \beta }q^{\beta \gamma }q^{\gamma \alpha  
}+\frac{u}{12}  
\sum _{ \alpha ,\beta }q^4_{ \alpha \beta }\right ],\\  
{{\cal L}}^{(1)}&=&\sqrt N\sum _{\alpha ,\beta }\Psi _{\vec  
{p}=0}^{  
\alpha \beta } \left [\tau q_{ \alpha \beta }+\frac{w}{2}  
(q^2)_{ \alpha \beta }+\frac{u}{3}q^3_{ \alpha \beta }\right  
],\\  
{{{\cal L}}}^{(2)}&=&-\frac{1}{2}\sum _{  
\alpha <  \beta,  
\gamma <  \delta  
} \sum _{\vec {p} }\Psi _{\vec {p} }^{ \alpha \beta  
}\left (\widetilde {G}^{-1}(\vec {p} )\right )_{ \alpha  
\beta,\gamma \delta  }\Psi _{-\vec {p} }^{\gamma \delta} \quad  
,  
\end{eqnarray}  
where  $\widetilde G^{-1}$ is the inverse of the free  
propagator:  
\begin{eqnarray}  
\left (\widetilde {G} ^{-1} (\vec {p} )\right )_{ \alpha  
\beta,\gamma \delta  }&=&(p^2\rho ^2a^2-2\tau -2uq^2_{ \alpha  
\beta  
})(\delta _{\alpha \gamma }^{Kr}\delta _{\beta \delta  
}^{Kr}+\delta  
_{\alpha \delta }^{Kr}\delta _{\beta \gamma }^{Kr})- \nonumber  
\\  
&&- w(\delta _{\alpha \gamma }^{Kr}q_{\beta \delta }+\delta  
_{\alpha  
\delta }^{Kr}q_{\beta \gamma }+\delta _{\beta \gamma  
}^{Kr}q_{\alpha \delta }+\delta _{\beta \delta }^{Kr}  
q_{\alpha \gamma})\quad ,\\  
{{\cal L}}^{(3)}&=&\frac{1}{\sqrt N}\left \{\frac{w}{6}\sum _{  
\alpha, \beta,\gamma  }\sum _{\vec {p_i}}\Psi _{\vec {p_1}}^{  
\alpha\beta }\Psi _{\vec {p_2}}^{\beta \gamma }\Psi _{\vec  
{p_3}}^{\gamma  
\alpha }+\frac{u}{3}\sum _{\alpha ,\beta }q_{ \alpha \beta  
}\sum  
_{\vec {p_i}} \Psi _{\vec {p_1}} ^{ \alpha \beta }\Psi _{\vec  
{p_2}}  
^{ \alpha \beta }\Psi _{\vec {p_3}} ^{ \alpha \beta }\right \}
\end{eqnarray}  
and finally  
\begin{eqnarray}  
{{\cal L}}^{(4)}&=&\frac{u}{12N}\sum _{ \alpha, \beta}\sum  
_{\vec  
{p_i}}\Psi _{\vec {p_1}}^{ \alpha \beta }\Psi _{\vec {p_2}}^{  
\alpha  
\beta }\Psi _{\vec {p_3}} ^{ \alpha \beta }\Psi _{\vec {p_4}}  
^{  
\alpha \beta }\quad .  
\end{eqnarray}  
  
We shall regard ${{\cal L}}^{(2)}$ as the bare Lagrangean and  
$${{\cal L}}^{(I)}={{\cal L}}^{(1)}+{{\cal L}}^{(3)}+{{\cal  
L}}^{(4)}$$  
as the interaction.  
  
Statistical averages are calculated with the weight $\sim {{\rm  
e}}^{{{\cal L}}}$. For example, the  expectation value of the  
fluctuation to first order in ${{\cal L}}^{(I)}$ works out as  
\begin{eqnarray}  
\!\!\!\!\!\!\!\!\!\!\!  
<\Psi _{\vec {p} }^{ \alpha \beta }> &=&\frac{\displaystyle  
\int  
[{{\rm d}}\Psi ]\Psi _{\vec {p} }^{ \alpha \beta }{{\rm  
e}}^{{{\cal  
L}}}}{  
\int [{{\rm d}}\Psi ]{{\rm e}}^{{\cal L}}}  
=\frac{\displaystyle \int [{{\rm d}}\Psi ]\Psi _{\vec {p} }^{  
\alpha \beta }{{\rm e}}^{{{\cal L}}^{(2)}}(1+{{\cal L}}^{(I)}+  
\ldots )}{  
\int [{{\rm d}}\Psi ]{{\rm e}}^{{{\cal L}}^{(2)}}(1+{{\cal  
L}}^{(I)}+ \ldots )}=  
<\Psi _{\vec {p} }^{ \alpha \beta }{{\cal  
L}}^{(I)}>_{(0)}+\ldots  
\end{eqnarray}  
where $<\ldots>_{(0)}$ is the average with the weight  ${{\rm  
e}}^{{{\cal L}}^{(2)}}$  and $<\Psi _{\vec {p} }^{ \alpha \beta  
}>_{(0)}=0$  has been used.  
  
By a Wick  decomposition Eq.~12 can further be written as:  
\begin{eqnarray}  
<\Psi _{\vec {p} }^{ \alpha \beta }>&=&\sqrt N\delta _{\vec {p}  
,0}^{Kr}\sum _{ \alpha ', \beta ' }\widetilde {G} _{ \alpha  
\beta  
,\alpha '\beta '}( \vec {p}=0)\left \{ \eta_{ \alpha ',\beta  
'}+\frac{w}{2N} \sum _{\vec {p} }\sum _{\gamma '\neq  
\alpha '\beta '}\widetilde {G} _{\alpha '\gamma ',\beta '\gamma  
'}(\vec {p})+\right .  
\nonumber \\  
&&+\left . \frac{u}{N}  
 \sum _{\vec {p} }q_{ \alpha '\beta '}\widetilde {G} _{\alpha  
'\beta ',\alpha '\beta '}(\vec {p})\right \}+\ldots \quad .  
\end{eqnarray}  
  
%In graphical terms  
%  
%\vspace*{-7mm}  
%\begin{center}  
%\begin{minipage}{10 cm}  
%\psfig{figure=t.eps}  
%\end{minipage}  
%\end{center}  
  
The quantity $\eta$ in Eq.~13  is  
\begin{eqnarray}  
\eta _{\alpha '\beta '}=\frac{1}{2}\left [2\tau q_{\alpha  
'\beta  
'}+ w(q^2)_{\alpha '\beta '}+\frac{2u}{3}q^3_{\alpha '\beta  
'}\right ].\end{eqnarray}  
  
The expectation value of the fluctuation must vanish, so we  
have to impose the condition  
\begin{eqnarray}  
\eta _{ \alpha \beta }+\frac{w}{2N}\sum _{\vec {p} }\sum  
_{\gamma \neq \alpha ,\beta }\widetilde {G} _{\alpha \gamma  
,\beta\gamma }(\vec {p} )+\frac{u}{N}\sum_{{\bf p}}  
\widetilde {G} _{\alpha \beta,\alpha\beta }(\vec {p} )  
q_{\alpha\beta}+\cdots=0\quad .\end{eqnarray}  
  
The propagator $\widetilde {G}$, as defined in Eq.~9, depends on  
the  
exact value  
of the order parameter, so (15) is, in principle, a  
self-consistent  
equation for $q_{\alpha \beta}$. For a long-ranged  
interaction $(z \gg 1)$,  
however, the loop corrections are small, as can be seen by  
changing the scale of the momentum so as to set the UV cutoff  
at $|{{\bf p}}|=1$:  
\begin{eqnarray}  
\frac{1}{N}\sum _{|\vec {p} |<1/\rho a}\ldots  
=\frac{a^d}{(2\pi)^d}\int _{|\vec {p} |<1/\rho a}{{\rm d}}^d\vec {p}  
\ldots =\frac{1}{z} \frac{1}{(2\pi)^d}\int _{|\vec {p} |<1}  
{{\rm d}}^d\vec {p}\ldots\quad .  
\end{eqnarray}  
  
If we call $G(\vec {p} ) $ the propagator $\widetilde {G} $  
with  
$\rho a$ absorbed into the  
momentum  
\begin{eqnarray}  
\left ( {G} ^{-1} (\vec {p} )\right )_{\alpha \beta ,\gamma  
\delta  
}&=&(p^2-2\tau -2uq^2_{ \alpha \beta })(\delta _{\alpha \gamma  
}^{Kr}\delta _{\beta \delta }^{Kr}+\delta _{\alpha \delta  
}^{Kr}\delta _{\beta \gamma }^{Kr})- \nonumber \\  
&-&w(\delta _{\alpha \gamma }^{Kr}q_{\beta \delta }+\delta  
_{\alpha  
\delta }^{Kr}q_{\beta \gamma }+\delta _{\beta \gamma  
}^{Kr}q_{\alpha \delta }+\delta _{\beta \delta  
}^{Kr}q_{\alpha \gamma}) \equiv(p^2 {\underline {\underline  
{\bf 1}}}+ {\underline {\underline{\bf M}}}^{(0)})_{\alpha  
\beta ,\gamma \delta }  
\end{eqnarray}  
then the equation of state (15) reads as  
\begin{eqnarray}  
&&2\tau q_{ \alpha \beta }+w(q^2)_{ \alpha \beta  
}+\frac{2u}{3}q^3_{  
\alpha \beta }+\nonumber \\ &&+  
\frac{1}{z}\frac{1}{(2\pi)^d}\int _{|\vec {p}|<1}{{\rm  
d}}^d\vec {p}\left [w\sum _{\gamma \neq\alpha ,\beta }G_{  
\alpha\gamma ,\beta \gamma }(\vec {p})+2uq_{ \alpha \beta }  
G_{ \alpha\beta  
,\alpha \beta }(\vec {p})\right ]+\cdots =0 \end{eqnarray}  
where the corrections are of ${{\cal O}}$($1/z^2$).  
  
Other physical quantities work out similarly, so by expanding  
in ${{\cal L}}^{(I)}$ we generate a series in powers of $1/z$  
with $  
G$ as the bare propagator. Before going into the analysis of  
the  
loop corrections, however, we have to solve the problem at zeroth order.  
  
\section {Mean field}  
  
Mean field theory is recovered in the present framework by  
letting the range of interaction go to infinity,  $z \to  
\infty$.  
In this limit all fluctuations vanish, $\Psi _{\vec {p}}  
\equiv 0$, and the free energy of the system is just given by  
 the constant in the Lagrangean:  
\begin{eqnarray}  
F=-T\lim_{n\to 0}\frac{1}{n}{{\cal L}}^{(0)}(q_{ \alpha \beta  
})  
\end{eqnarray}  
evaluated at the solution of  
\begin{eqnarray}  
2\tau q_{ \alpha \beta }+w(q^2)_{ \alpha \beta  
}+\frac{2u}{3}q^3_{  
\alpha \beta }=0\quad .\end{eqnarray}

Because of the replica limit, $n\to 0$, involved in the  
formalism,  
one has to solve Eq.~20 in  the space of $0\times 0$ matrices  
which  
implies  
that one is able to parametrize the matrix $q_{ \alpha \beta }$  
in  
such a way as to permit the limit $n\to 0$ to be taken. The  
obvious  
Ansatz, due to Sherrington and Kirkpatrick [2], is to assume  
that  
$q_{ \alpha \beta }$   does not depend on the replica indices,  
i.e.  
$$q_{ \alpha \beta }^{SK}=q(1-\delta _{ \alpha \beta  
}^{Kr})\quad .$$  
  
In the limit  $n\to 0$ Eq.~20 would then become  
\setcounter{equation}{19}\primeq \begin{eqnarray}  
2\tau q-2wq^2+\frac{2u}{3}q^3=0\quad .  
\end{eqnarray} \normeq  
However, the assumption of replica symmetry  
led to paradoxical results at low temperatures, and  
was subsequently shown to be unstable against replica symmetry  
breaking fluctuations by de Almeida and Thouless [31].  
  
After some unsuccessful attempts by various groups, the solution  
which, at least within the context of mean field theory, is  
now generally accepted was found by Parisi [1]. Parisi's  
solution is based on a hierarchical replica symmetry breaking  
pattern which reflects the ultrametric organization of  
equilibrium states in the long range spin glass [37].  
  
We do not need to go into the details of the Parisi solution  
here, so we discuss it only to the extent necessary for fixing  
notation.  
  
We call the sizes of the Parisi  blocks $p_r$, $r=1,2,\ldots R$  
with $R$, the number of RSB steps, going to infinity at the  
end.  
For  
the sake of uniformity of notation it is convenient to add  
$p_0\equiv n$ and $p_{R+1}\equiv 1$  to the two  
ends of the series $p_r$. The value of the order parameter on  
the  
$r^{\rm th}$ level of hierarchy will be called $q_r$,  
$r=1,2,\ldots R$. Upon analytic continuation in $n$ the series  
$p_r$ becomes monotonically increasing.  
  
A useful concept we shall make frequent use of in the following  
is that of the overlap between replica indices: $\alpha \cap  
\beta  
$ is essentially the inverse of $q_{ \alpha \beta }$, that is  
\begin{eqnarray}  
\alpha \cap \beta =r \quad \mbox{{\rm if the corresponding}}  
\quad  
q_{ \alpha \beta }=q_r\quad .\end{eqnarray}  
Accordingly, the allowed values for  $\alpha \cap \beta $ range  
from $\alpha \cap \beta =0$  
(corresponding to the outermost region in Parisi's pattern) to  
$\alpha \cap \beta =R$  (in the innermost block). By extension,  
we  
add $\alpha \cap \beta =R+1$,  
corresponding to the diagonal, $\alpha =\beta $.  
  
The overlap $\alpha \cap \beta $ is a kind of  hierarchical  
codistance between replicas $\alpha $ and $\beta $. The metric  
generated by the overlaps is ultrametric:  
whichever way we choose three replicas $\alpha$, $\beta$,  
$\gamma$,  
either all three of  
their overlaps are the same $\alpha \cap \beta =\alpha \cap  
\gamma  
=\beta \cap \gamma $, or one  (say $\alpha \cap \beta $)  is  
larger  
than the other two, but then these are equal ($\alpha \cap  
\beta  
>\alpha \cap \gamma =\beta \cap \gamma $).  
  
It is evident that any quantity  $f$  constructed of the $q$'s  
and  
depending on two replica indices (like $(q^2)_{ \alpha \beta  
}=\sum  
_{\gamma }q_{\alpha \gamma }q_{\gamma \beta }$, e.g.)  depends  
only  
on their overlap: $f_{\alpha \beta }=f(\alpha \cap \beta )$.  
Furthermore, any quantity    $f_{\alpha \beta \gamma }$  
depending  
on three replicas depends only  on the three overlaps, and  
since of  
these at most two can be different,   $f_{\alpha \beta \gamma  
}$  
is, in fact, a function of only two variables, e.g.\ $\alpha  
\cap  
\beta $  and the larger of the other two:  
\begin{eqnarray}  
f_{\alpha \beta \gamma }=f(\alpha \cap \beta ,\max\{\alpha \cap  
\gamma ,\beta \cap \gamma \})\quad .  
\end{eqnarray}  
  
In the following we will also have to deal with quantities  
depending on two pairs of replicas, like the propagator  $G_{  
\alpha \beta ,\gamma \delta }$, for example.  
Ultrametrics implies that of the six possible overlaps between  
$\alpha$, $\beta$, $\gamma$, $\delta$ at  most three can be  
 different, which corresponds to the  
elementary geometric fact that a tetrahedron built of  
equilateral and isosceles faces can only have three different  
 edges. For these  
4-replica quantities we proposed the following parametrization  
\begin{eqnarray}  
G_{\alpha\beta , \gamma\delta}=G^{\alpha\cap\beta,\gamma\cap  
\delta}_{\max\{\alpha\cap \gamma,\alpha\cap\delta\},  
\max\{\beta\cap \gamma,\beta\cap\delta\}}
\end{eqnarray}  
in [28]. This parametrization is redundant in that, according to  
what has just been said, of the four variables displayed in  
Eq.~23 at  
least two must coincide, on the other hand it has the merit of  
being  
symmetric.  
  
We shall keep to this parametrization throughout most of the  
paper,  
in order to ensure consistency with previously published  
material. In Sec.~7,  
however, the redundancy would mask an important property, so  
there we  
will make use of the following observation: The only case when  
the two lower  
indices in Eq.~23 carry independent information is when the two  
upper indices  
coincide {\it and} both lower indices are larger than the  
common value of  
the upper ones. In all other cases the smaller of the two lower  
indices is a  
dummy variable which we may drop and keep only  
$\max\{\alpha\cap\gamma,  
\alpha \cap\delta,\beta \cap \gamma,\beta\cap\delta\}$ as the  
sole lower  
index.  
  
A frequently used symbol will be  
\begin{eqnarray}  
\delta _r=p_r-p_{r+1}\quad .  
\end{eqnarray}  
  
In the limit $R\to \infty$, $q_r$ goes over into a continuous,  
monotonically increasing function $q(x)$, which turns out to have a  
breakpoint  $x_1$,  
beyond which it is constant. Although for large $R$ the precise  
choice of the block sizes is largely immaterial, it is  
convenient to arrange the $p_r$'s so that they  fill the  
interval  ($0,x_1$)  with $\delta _r$, $r=0,1,2,\ldots , R-1$,  
becoming infinitesimal, of ${{\cal O}}$($1/R$), and the  
last one $\delta _R=p_R-p_{R+1}=p_R-1\to(x_1-1)$ staying  
finite.  
  
Returning now to the equation of state (20), under the Parisi  
parametrization it becomes:  
\begin{eqnarray}  
2\tau q_r+w\left [\sum _{t=0}^r\delta _tq^2_t-  
p_{r+1}q^2_r+2q_r\sum  
_{t=r+1}^R\delta _tq_t\right  
]+\frac{2u}{3}q^3_r=0\quad ,\end{eqnarray}  
or, for $R\to \infty$:  
\begin{eqnarray}  
2\tau q(x)-w\left [\int _0^x{{\rm d}}tq^2(t)+xq^2(x)+2q(x)\int  
_x^1{{\rm d}}tq(t)\right ]+\frac{2u}{3}q^3(x)=0\quad  
.\end{eqnarray}  
This is easily solved by repeated differentiation yielding the  
well-known result [1]  
\begin{eqnarray}  
q(x)&=&\left \{  
\begin{array}{ll}  
\displaystyle \frac{w}{2u}x, & x<x_1 \\ & \\  
\displaystyle \frac{w}{2u}x_1\equiv q_1, & x_1< x< 1 \\  
\end{array}\right.\quad ,\end{eqnarray}  
with the breakpoint $x_1$ to be determined from the condition  
on $q_1$  
\begin{eqnarray}  
\tau -wq_1+uq_1^2=0\quad .  
\end{eqnarray}  
  
To leading order in $\tau$, $q_1=\tau /w $   which could have  
been read off from  
Eq.~26 directly, by noting that for $x_1\sim q_1\sim\tau $  
all the terms in Eq.~26 are of ${{\cal O}}$($\tau ^3$) except the  
first and the one  
coming from the upper end of the $\int _x^1$ integral. This  
type of  
approximation, consisting in  
dropping all replica integrals and keeping only the  
contribution coming from the vicinity of $x=1$, will be used  
frequently in the following and will be referred to as the  
``innermost block approximation".  
  
The solution given in Eq.~27 is valid as long as $\tau >0$  
($T<T_c$).  
With $T\to T_c$, $q(x)\to 0$ and one enters the paramagnetic  
phase.  
  
The physical meaning of the order parameter  $q(x)$ depending  
on a  
continuous variable had remained a mystery until Parisi showed  
[38]  
that the derivative of its inverse  
\begin{eqnarray}  
\frac{{{\rm d}}x}{{{\rm d}}q}=P(q)  
\end{eqnarray}  
is nothing but the probability distribution of the magnetic  
overlaps  $q_{ab}$ between the equilibrium states $a$ and $b$  
of  
the system:  
\begin{eqnarray}  
P(q)&=&\overline {<\delta (q-q_{ab})>}\\  
q_{ab}&=&\frac{1}{N}\sum _i<s_i>_a<s_i>_b \quad . 
\end{eqnarray}  
(In  Eq.~30 $<\ldots >$ is the thermal average and the overbar  
is  
the average over the random couplings.)

\section{Instability of the replica symmetric solution and mass  
scales}  
  
In order to go one step beyond mean field theory, we turn now  
to  
the study of the quadratic part ${{\cal L}}^{(2)}$ of the  
Lagrangean which describes free (Gaussian) fluctuations of  the  
order parameter. The main task is to diagonalize the quadratic  
form  
in Eq.~8, i.e.\ to invert (17) and obtain the free propagator $G$.  
The  
spectrum of free fluctuations (the singularities of $G$) will  
also  
provide a stability test for the mean field solution found  
above.  
Multiplying Eq.~17 by $G$ we get the following set of equations  
for  
the propagator components:  
\begin{eqnarray}  
 (p^2-2\tau -2uq^2_{ \gamma \delta })G_{ \alpha \beta ,\gamma  
\delta }-w\sum _{\mu \neq \gamma ,\delta }q_{\mu \delta }G_{  
\alpha \beta ,\gamma \mu }  
-w\sum _{\mu \neq\gamma ,\delta }q_{\mu \gamma }G_{ \alpha  
\beta,\mu \delta }=\delta _{\alpha \gamma }^{Kr}\delta _{\beta  
\delta}^{Kr}\quad ,&&\nonumber\\  
 \alpha < \beta ,\quad \gamma < \delta \quad . &&  
\end{eqnarray}  
  
For a first orientation, let us work out the  
solution in the replica symmetric case. Then $G_{ \alpha \beta  
,\gamma \delta }$  will have only three different components  
\begin{eqnarray}  
{\displaystyle \begin{array}{lll}  
{\displaystyle G_{ \alpha \beta ,\alpha \beta }=G_1\quad ,} &  
\alpha  
\neq  
\beta & \\  
G_{ \alpha \beta ,\alpha \gamma }=G_2\quad , &\alpha ,\beta  
,\gamma  
& \mbox{{all different,}}\\  
G_{ \alpha \beta ,\gamma \delta }=G_3\quad , & \alpha ,\beta  
,\gamma  
,\delta  & \mbox{{all different,}}\\  
\end{array}}  
\end{eqnarray}  
satisfying  the simple set of equations  
\begin{eqnarray}  
\left (p^2-2wq-\frac{4u}{3}q^2\right )G_1  
+4wqG_2&=&1\nonumber \\  
\left (p^2-\frac{4u}{3}q^2\right )G_2-wqG_1+3wqG_3&=&0 \\  
\left (p^2+6wq-\frac{4u}{3}q^2\right )G_3-4wqG_2&=&0\nonumber  
\end{eqnarray}  
where we have used Eq.~20' and put $n=0$.  
The solutions to (34), first written up by Pytte and Rudnick [39],
\begin{eqnarray}  
G_1&=&\frac{1}{p^2-\frac{4u}{3}q^2}\left  
(1+\frac{2wq}{p^2+2wq-\frac{4u}{3}q^2}+  
\frac{4w^2q^2}{(p^2+2wq-\frac{4u}{3}q^2)^2}\right )\nonumber \\  
G_2&=&\frac{1}{p^2-\frac{4u}{3}q^2}\left  
(\frac{1}{2}\frac{2wq}{p^2+2wq-\frac{4u}{3}q^2}
+\frac{4w^2q^2}{(p^2+2wq-\frac{4u}{3}q^2)^2} 
\right )\\  
G_3&=&\frac{1}{p^2-\frac{4u}{3}q^2}
\frac{4w^2q^2}{(p^2+2wq-\frac{4u}{3}q^2)^2}  
\nonumber\end{eqnarray}  
display the AT instability [31] in  
that the first factors have a pole on the positive $p^2$ axis.  
Note  
that the instability is due entirely to the quartic  
interaction: if we had set $u=0$ in the Lagrangean (2), we  
would  
have found an acceptable pole structure in $G$, at least on the  
level of Gaussian approximation. Also note that there are  
two ``mass-scales" displayed by Eq.~35: the unstable mode has a  
``mass" of order $\tau$  (the small mass), while the  
stable  mass is ${{\cal O}}$($\tau^{1/2} $) (the ``large" mass).  
  
\section{The large momentum behaviour}  
  
When the assumption of replica symmetry is abandoned the  
solution of Eq.~32 immediately becomes very difficult. It is,  
therefore, important to  realize that there are instances when  
we do not need to get seriously involved  with the intricacies  
of the  
symmetry broken solution. In  
sufficiently high dimensions the loop corrections to the  
various physical quantities depend upon the short distance  
behaviour of the propagators only, even in the vicinity of $T_c$, and  
for  
such large values of the wavevector (around the upper cutoff,  
i.e.\ much larger than either of the characteristic masses) one  
should be able to get a solution to Eq.~32 by simply expanding  
$G$  
for large $p^2$, without having to introduce any particular  
parametrization for the order parameter.  
  
We have to keep in mind, of course, that on passing from the  
original model (1) to the effective Lagrangean (2) we have dropped  
all microscopic details, so we can certainly not expect to  
retrieve the precise  
short distance behaviour from the large-$p$ expansion of  
Eq.~32. The qualitative behaviour will, however, be correct and the  
results will later allow us to illustrate some important points  
about  
how the  renormalization of the coupling constants takes place  
in high dimensions, and also about the characteristic  
dimensions of the model.  
  
In order to implement the program sketched above, we need to  
decompose (32)  so as to separate  terms containing propagator  
components of the first, second, and third  kind  
($G_{\alpha\beta ,\alpha \beta}$, $G_{\alpha\beta ,\alpha  
\gamma}$,  
 and $G_{\alpha\beta ,\gamma\delta}$ with $\alpha , \beta ,  
\gamma ,  
 \delta$ all different, respectively). Dividing  
through by  
 $(p^2-2\tau -2uq^2_{ \alpha\beta })$  
we also wish to write Eq.~32 in the form of a set of  
Dyson equations. Thus we have:  
\begin{eqnarray}  
G_{ \alpha \beta ,\alpha \beta }&=&G^{(0)}_{ \alpha \beta  
,\alpha  
\beta }+w\sum _{\omega \neq\alpha ,\beta }G^{(0)}_{ \alpha  
\beta  
,\alpha \beta }q_{\omega \beta }G_{ \alpha \beta ,\alpha \omega  
}  
+w\sum _{\omega \neq\alpha ,\beta }G^{(0)}_{ \alpha  
\beta ,\alpha \beta }q_{\omega \alpha  }G_{ \alpha \beta  
,\omega  
\beta } \\  
% (37)  
%  
G_{ \alpha \beta ,\alpha\gamma  }&=&wG^{(0)}_{\alpha\gamma  
,\alpha\gamma }q_{\beta\gamma}G_{\alpha \beta ,\alpha \beta}  
+w\sum_{\omega \neq \alpha,\beta,\gamma}G^{(0)}_{\alpha\gamma  
,\alpha\gamma }q_{\omega\gamma}G_{\alpha \beta ,\alpha \omega}  
+\nonumber\\  
&&+w\sum_{\omega \neq \alpha,\beta,\gamma}G^{(0)}_{\alpha\gamma  
,\alpha\gamma }q_{\omega\alpha}G_{\alpha \beta ,\omega\gamma}+  
wG^{(0)}_{\alpha\gamma ,\alpha\gamma }q_{\alpha\beta}  
G_{\alpha \beta , \beta\gamma}\\  
%  
% (38)  
%  
G_{ \alpha \beta ,\gamma \delta }&=&wG^{(0)}_{\gamma \delta  
,\gamma  
\delta }\left (q_{\alpha \delta }G_{\alpha \beta ,\alpha \gamma  
}+q_{\beta \delta }G_{ \alpha \beta ,\beta \gamma }+q_{\alpha  
\gamma }G_{ \alpha \beta ,\alpha \delta }+q_{\beta \gamma }G_{  
\alpha \beta ,\beta \delta }\right )+\nonumber \\  
&&  
+w\sum _{\omega \neq\alpha ,\beta,\gamma ,\delta  
}G^{(0)}_{\gamma \delta ,\gamma \delta }q_{\omega \delta }G_{  
\alpha \beta ,\gamma \omega }+w\sum _{\omega \neq\alpha  
,\beta,\gamma ,\delta  }G^{(0)}_{\gamma \delta ,\gamma \delta  
}q_{\omega\gamma }G_{ \alpha \beta ,\omega \delta }  
\end{eqnarray}  
where  
\begin{eqnarray}  
G^{(0)}_{ \alpha \beta ,\alpha \beta }=\frac{1}{p^2-2\tau -  
2uq^2_{  
\alpha \beta }}\quad .\end{eqnarray}  
These equations are solved iteratively by noticing that for  
``large $p^2$" (i.e.\ for $p^2\gg \tau ,q$)  the only term that  
survives is $G^{(0)}$ in Eq.~36, so  
\begin{eqnarray}  
G_{ \alpha \beta ,\alpha \beta }=\frac{1}{p^2}+\cdots.  
\end{eqnarray}  
This is now substituted into Eq.~37 to yield  
\begin{eqnarray}  
G_{ \alpha \beta ,\alpha \gamma }=\frac{wq_{\beta \gamma  
}}{p^4}+\cdots,  
\end{eqnarray}  
and so forth. The results one obtains after the first few  
iterations are the following:  
\begin{eqnarray}  
G_{ \alpha \beta ,\alpha \beta }&=&\frac{1}{p^2}+\frac{2\tau  
+2uq^2_{ \alpha \beta }}{p^4}+\frac{1}{p^6}\left \{(2\tau  
+2uq^2_{  
\alpha \beta })^2+w^2\left ((q^2)_{ \alpha \alpha  
}+(q^2)_{\beta  
\beta }-2q^2_{ \alpha \beta }\right )\right \}+\nonumber \\  
&&+\frac{1}{p^8}\left \{(2\tau +2uq^2_{ \alpha \beta  
})^3+6w^2\tau  
\left ((q^2)_{ \alpha \alpha  }+(q^2)_{\beta \beta }-2q^2_{  
\alpha  
\beta }\right )+\right. \nonumber \\  
&&+ 4uw^2\sum _{\omega }q^2_{\alpha \omega} q^2_{\omega \beta  
}+  
+ 4uw^2q^2_{ \alpha \beta }\left ( (q^2)_{ \alpha \alpha  
}+(q^2)_{\beta \beta }-2q^2_{ \alpha \beta }\right )+\nonumber  
\\  
&&+\left. w^3\left( (q^3)_{ \alpha \alpha  }+(q^3)_{\beta \beta  
}  
-2q_{ \alpha \beta }(q^2)_{ \alpha \beta }\right )\right  
\}+\cdots\\  
% G_{\alpha \gamma ,\beta \gamma }&=&\frac{wq_{ \alpha \beta  
%}}{p^4}+\frac{1}{p^6}\left [4w\tau q_{ \alpha \beta  
%}+w^2(q^2)_{  
%\alpha \beta }+2uwq_{ \alpha \beta }(q^2_{\alpha \gamma  
%}+q^2_{\beta \gamma })\right ]+\nonumber \\  
%&&+\frac{1}{p^8}\left [12w\tau ^2q_{ \alpha \beta }+6\tau  
%w^2(q^2)_{ \alpha \beta }+w^3(3q_{ \alpha \beta }(q^2)_{\gamma  
%\gamma }-4q_{ \alpha \beta }(q^2)_{\beta \gamma  
%}(q^2_{\alpha \gamma}+q^2_{\beta \gamma})+\right.\nonumber\\  
%&&\left. +(q^3)_{ \alpha \beta }+q_{\alpha  
%\gamma }  
%(q^2)_{\beta \gamma }+q_{\beta \gamma }(q^2)_{\alpha \gamma } )  
%\right ]+\cdots \\  
G_{\alpha \gamma ,\beta \gamma }&=&\frac{wq_{ \alpha \beta  
}}{p^4}+\frac{1}{p^6}\left \{4w\tau q_{ \alpha \beta  
}+w^2(q^2)_{  
\alpha \beta }+2uwq_{ \alpha \beta }(q^2_{\alpha \gamma  
}+q^2_{\beta \gamma })\right \}+\nonumber \\  
&&+\frac{1}{p^8}\left \{12w\tau ^2q_{ \alpha \beta }+6\tau  
w^2(q^2)_{ \alpha \beta }+w^3\left(3q_{ \alpha \beta }(q^2)_{\gamma  
\gamma }-4q_{ \alpha \beta }
(q^2_{\alpha \gamma}+q^2_{\beta \gamma})+\right.\right.\nonumber\\  
&&\left.+(q^3)_{ \alpha \beta }+q_{\alpha  
\gamma }  
(q^2)_{\beta \gamma }+q_{\beta \gamma }(q^2)_{\alpha \gamma }\right )  
+12uw\tau q_{\alpha\beta}(q^2_{\alpha \gamma}+
q^2_{\beta \gamma})+ \nonumber\\  
&&+2uw^2\left((q^2)_{\alpha\beta}(q^2_{\alpha \gamma}+
q^2_{\beta \gamma})+q^2_{\alpha\beta}q_{\alpha\gamma}q_{\beta\gamma}
+\sum _{\omega }q^2_{\gamma \omega} q_{\alpha\omega}q_{\beta\omega}  
\right)+\nonumber\\
&&+\left. 4u^2w q_{\alpha\beta} (q_{\alpha\gamma}^4+q_{\beta\gamma}^4+
q_{\alpha\gamma}^2q_{\beta\gamma}^2)\right\}+\cdots \\
G_{ \alpha \beta ,\gamma \delta }&=&\frac{2w^2}{p^6}(q_{\alpha  
\delta }q_{\beta \gamma }+q_{\alpha \gamma }q_{\beta \delta  
})+\cdots\quad .  
\end{eqnarray}  
  
We will use these expressions in Sec.~10 to calculate short  
range corrections to the equ-\linebreak ation of state and to  
the  
propagators in high ($d>8$) dimensions.

\section{The near infrared region}  
  
Now we wish to probe deeper into the structure of the  
propagator  
and investigate its behaviour in the long, but not extremely  
long wavelength region. More  precisely, what we mean is that the  
momentum may be comparable to the large mass  
but it remains much larger than the small mass scale:  
\begin{eqnarray}  
p^2 \gg \tau ^2  
\end{eqnarray}  
We call this region the near infrared region.  
  
In order to solve the problem for momenta that are comparable  
to  
the large mass ($p^2\sim\tau $)  we have to construct  
the equations for the various components of the propagator 
explicitly,  by  
writing out  Eq.~32 under the parametrization  (23) and  
letting $R$ go to infinity at the end. The set of integral  
equations  
obtained this way was first published in [28]; we display them 
here for the sake of completeness.  
\begin{eqnarray}  
&&\left (p^2-2\tau-2uq^2(x) \right )G^{x,x}_{1,1}+2w\int  
_0^x {{\rm d}}tq(t)G^{x,t} _{1,x}+\nonumber\\  
&&\qquad  
+ 2wxq(x)G^{x,x}_{1,x}+2w\int ^1_x{{\rm d}}tq(t)  
G^{x,x}_{1,t}+2wq(x)\int _x^1{{\rm d}}tG_{1,x}^{x,t}=1\quad ,\\[2mm]
%  
% 2  
%  
&&\left (p^2+uq^2(z)+uq_1^2-2uq^2(x) \right )  
G^{x,x}_{z,1}+w\int_0^x{{\rm d}}tq(t)G_{z,x}^{x,t}+  
\nonumber \\  
&&\qquad +w\int _x^1{{\rm d}}tq(t)G^{x,x}_{z,t}-  
wq_1G^{x,x}_{z,1}+w\int _x^z{{\rm d}}tq(t)G^  
{x,x}_{1,t}+\nonumber\\  
&&\qquad + wq(x)\left \{xG^{x,x}_{z,x} +  
\int _x^1{{\rm d}}tG^{x,t}_{z,x}+zG^{x,z}_{z,x}+\int  
_z^1{{\rm d}}tG^{x,z}_{t,x}-G^{x,z}_{1,x}\right  
\}+w\int _0^x{{\rm d}}tq(t)G_{1,x}^{x,t}+\nonumber\\  
&&\qquad +wq(z)\left \{\int _z^1{{\rm  
d}}tG_{1,t}^{x,x}-G_{1,1}^{x,x}\right  
\}+wq(x)\left \{xG_{1,x}^{x,x}+\int _x^1{{\rm d}}t  
G_{1,x}^{x,t}\right \}=0\quad \!\!\!\! , \!\!\!\!\quad x\le z<1,\\[2mm]
%  
% 3  
%  
&&\left (p^2+uq_1^2-uq^2(y)\right )G^{x,y}_{1,x}+w\int  
_0^y{{\rm d}}tq(t)\left (G^{x,t}_{1,x}+  
G_{y,x}^{x,t}\right ) + w\int _y^1{{\rm d}}tq(t)  
G_{t,x}^{x,y}+\nonumber \\  
&&\qquad  +wq(y)\left \{yG_{y,x}^{x,y}+ \int _y^1{{\rm  
d}}tG_{y,x}^{x,t}+  
\int _y^1{{\rm d}}tG_{1,x}^{x,t}\right \}  
+\nonumber \\  
&&\qquad  
 +wq(x)\left \{xG_{y,x}^{x,x}+\int _x^1{{\rm d}}t  
G_{y,t}^{x,x}+xG_{1,x}^{x,x}+ \int _x^1{{\rm d}}tG_{1,t}^{x,x}-  
G_{y,1}^{x,x}-G_{1,1}^{x,x}\right \}-\nonumber\\
&&\qquad -wq_1G_{1,x}^{x,y}=0\quad , \qquad  
x\le y < 1\quad ,  
\\[2mm]
%\begin{array}{r}\displaystyle  
% +wq(x)\left \{xG_{y,x}^{x,x}+\int _x^1{{\rm d}}t  
%G_{y,t}^{x,x}+xG_{1,x}^{x,x}+ \int _x^1{{\rm d}}tG_{1,t}^{x,x}-  
%G_{y,1}^{x,x}-G_{1,1}^{x,x}\right \}-\\
%-wq_1G_{1,x}^{x,y}=0\quad , \qquad  
%\displaystyle x\le y < 1\quad ,  
%\end{array}\\  
%  
% 4  
%  
&&\left (p^2+uq_1^2-uq^2(y)\right )G_{1,x}^{x,y}+w\int  
_0^y{{\rm d}}tq(t)\left (G_{1,x}^{x,t}+  
G_{y,y}^{x,t}\right )+w\int _y^x{{\rm d}}tq(t)  
G_{t,t}^{x,y}+\nonumber\\  
&&\qquad +w\int _x^1{{\rm d}}tq(t)G_{t,x}^{x,y}+  
wq(y)\left \{\int _y^1G_{y,y}^{x,t}+  
yG _{y,y}^{x,y}+\int _y^1{{\rm d}}tG_{1,x}^{x,t}+  
\right.\nonumber\\  
&&\qquad \left. +x G_{1,x}  
^{x,x}+\int _x^1{{\rm d}}tG_{1,t}^{x,x} -  
G_{1,1}^{x,x}\right \}+\nonumber\\  
&&\qquad +wq(x)\left \{xG_{x,x}^{x,y}+\int _x^1{{\rm d}}t  
G_{x,t}  
^{x,y}-G_{x,1}^{x,y}\right \}-wq_1  
G_{1,x}^{x,y}=0\quad , \quad y \le x<1\quad , \\[2mm]
%  
% 5  
%  
&&\frac{1}{2}p^2G_{z,z}^{x,y}+w\int _0^y{{\rm d}}  
tq(t)G_{z,z}^{x,t}+wq(y)\int _y^1{{\rm d}}t  
G_{z,z}^{x,t}+ wq(z)\left \{ zG_{z,z}^{x,z}+\int _z^x{{\rm d}}  
tG_{t,t}^{x,z}+\right.\nonumber\\  
&&\qquad \left. +xG_{x,x}^{x,z}+2\int  
_x^1{{\rm d}}tG_{t,x}^{x,z}-2G_{1,x}^{x,z}\right  
\}=0\quad , \quad  z\le x,y <1\quad , \\[2mm]
%  
% 6  
%  
&&\frac{1}{2}p^2G_{z,x}^{x,y}+w\int _0^y{{\rm d}}  
tq(t)G_{z,x}^{x,t}+wq(y)\int _y^1{{\rm d}}t  
G_{z,x}^{x,t}+  
wq(z) \left \{zG_{z,x}^{x,z}+\int _z^1{{\rm d}}t  
G _{ t,x}^{x,z}-G_{1,x}^{x,z}\right \}+  
\nonumber \\  
&&\qquad +wq(x)\left \{x  
G_{z,x}^{x,x}+\int _x^1{{\rm d}}tG_{z,t}^{x,x}-  
G_{z,1}^{x,x}\right \}=0\quad , \qquad x \le z \le y <1\quad ,  
\\[2mm]
%  
% 7  
%  
&&\left (p^2+uq^2(z)-uq^2(y)\right )G_{z,z}^{x,y}+w\int  
_0^y{{\rm d}}tq(t)\left (G_{y,y}^{x,t}+  
G_{z,z}^{x,t}\right )+w\int _y^z{{\rm d}}tq(t)  
G_{t,t}^{x,y}+\nonumber \\  
&&\qquad +wq(y)\left \{\int_y^1{{\rm d}}t  
G_{y,y}^{x,t}+yG_{y,y}^{x,y}+\int _y^1{{\rm d}}t  
G_{z,z}^{x,t}+zG_{z,z}^{x,z}+  
\right.\nonumber\\  
&&\qquad +\left.\int _z^x{{\rm  
d}}tG_{t,t}^{x,z}+xG_{x,x}^{x,z}+  
2 \int _x^1{{\rm d}}tG_{t,x}^{x,z}-2G_{1,x}^{x,z}  
\right \}+\nonumber\\  
&&\qquad +wq(z)\left\{xG_{x,x}^{x,y}+\int _z^x{{\rm  
d}}tG_{t,t}^{x,y}+2\int _x^1{{\rm d}}t  
G_{t,x}^{x,y}-2G_{1,x}^{x,y}\right \}=0\quad \!\!  
,\!\!\!\!\quad y\le z  
\le x < 1\quad \!\!\!\! , \\[2mm]
%  
% 8  
%  
&&\left (p^2+uq^2(z)-uq^2(y)\right )G_{z,x}^{x,y}+w\int  
_0^y{{\rm d}}tq(t)\left (G_{z,x}^{x,t}+  
G_{y,x}^{x,t}\right )+w\int _y^z{{\rm d}}tq(t)  
G_{t,x}^{x,y}+\nonumber \\  
&&\qquad +wq(y)\left \{yG_{y,x}^{x,y}+\int_y^1{{\rm  
d}}tG_{y,x}^{x,t}  
+\int _z^1{{\rm d}}t\left (G_{z,x}^{x,t}+  
G_{t,x}^{x,z}\right )+zG_{z,x}^{x,z}  
+\right.  
\nonumber \\&&\qquad \left.  
+\int _y^z{{\rm d}}tG_{z,x}^{x,t} -G_{1,x}^{x,z}\right \}  
+wq(x)\left\{xG_{y,x}^{x,x}+\int _x^1{{\rm d}}t  
G_{y,t}^{x,x}+xG_{z,x}^{x,x}+\int _x^1{{\rm d}}t  
G_{z,t}^{x,x}-\right .\nonumber\\  
&&\qquad  
\left. -G_{y,1}^{x,x}-G_{z,1}^{x,x}\right \}+  
wq(z)\left \{\int _z^1{{\rm d}}tG_{t,x}^{x,y}-  
G_{1,x}^{x,y}  
\right \}=0\quad ,\qquad x\le y \le z  < 1\quad , \\[2mm]
%  
% 9  
%  
&&\left (p^2+uq^2(z)-uq^2(y)\right )G_{z,x}^{x,y}+w\int  
_0^y{{\rm d}}tq(t)\left (G_{y,y}^{x,t}+  
G_{z,x}^{x,t}\right )+w\int _y^x{{\rm d}}tq(t)  
G_{t,t}^{x,y}+\nonumber\\  
&&\qquad +w\int _x^z{{\rm d}}tq(t)G_{t,x}^{x,y}+wq(y)  
\left \{\int _y^1{{\rm d}}tG_{y,y}^{x,t}+y  
G_{y,y}^{x,y}+\int_y^1{{\rm dt}}G_{z,x}^{x,t}+  
zG_{z,x}^{x,z}+\right.\nonumber\\  
&&\qquad+\int _z^1{{\rm d}}t  
G_{t,x}^{x,z}+xG_{z,x}^{x,x}+\left.\int _x^1{{\rm d}}t  
G_{z,t}^{x,x}-G_{1,x}^{x,z}-G_{z,1}^{x,x}\right  
\}+\nonumber\\  
&&\qquad +  
wq(x)\left \{xG_{x,x}^{x,y}+\int _x^1{{\rm d}}t  
G_{x,t}^{x,y}-G_{x,1}^{x,y}\right \}+\nonumber\\  
&&\qquad +wq(z)\left \{ \int  
_z^1{{\rm d}}tG_{t,x}^{x,y}-G_{1,x}^{x,y}\right  
\}=0\quad , \qquad y\le x \le z < 1\quad ,\\[2mm]
%  
% 10  
%  
&&\left (p^2+uq^2(z_1)+uq^2(z_2)-2uq^2(x)\right )  
G_{z_1,z_2}^{x,x}+\nonumber\\  
&&\qquad +A(z_1,z_2)+A(z_2,z_1)=0\quad , \quad  
x\le z_1,z_2 <1\quad ,\\  
\!\!\!\!\!\!\!\!\!\!\!\!\!\!\!\!\!\!\!\!\!\!  
{{\rm where}}\!\!\!\!\!\!\!  
\!\!\!\!\!\!\!\!\!  
&&\nonumber\\  
&&A(z_1,z_2)=w\int _0^x{{\rm d}}tq(t)  
G_{z_1,x}^{x,t}+w\int _x^{z_2}{{\rm d}}tq(t)  
G_{z_1,t}^{x,x}+\nonumber \\  
&&\qquad +wq(z_2) \left \{\int _{z_2} ^1{{\rm d}}t  
G_{z_1,t}^{x,x}- G_{z_1,1}^{x,x}\right \}+ wq(x) \left \{  
xG_{z_1,x}^{x,x}+\int _x^1  
G_{z_1,x}^{x,t}+z_1G_{z_1,x}^{x,z_1}+  
\right.\nonumber\\  
&&\qquad +\left. \int _{z_1}^1{{\rm d}}t  
G_{t,x}^{x,z_1}-G_{1,x}^{x,z_1}\right \},\quad x \le  
z_1,z_2 <1\quad .  
\end{eqnarray}

In Eqs.~46$-$56, the mean field equations (27), (28)
have been repeatedly used.
  
Note that Eqs.~47, 48 and 49 can  formally be obtained  
from  
Eqs.~55, 53 and
%\linebreak
54, respectively, (putting  
$z_1=z$,  
$z_2=1$ in Eq.~55,  and $z=1$ in Eq.~53 and 54)  
which is why we did not write them up explicitly in [28]. The  
reason for which we display them here is twofold. Firstly, it  
is not immediately obvious why Eq.~55 should contain Eq.~47,  
etc., since, although  $G_{z_1,z_2}^{x,y}$ is, in general, a  
continuous  
function of  its variables, this continuity does not apply when  
either of the  
lower variables takes the value 1. (Whenever this happens we  
have coinciding replica indices,  and there is no reason to  
expect e.g.\ $G_{ \alpha \beta ,\gamma \delta }$, with $\alpha  
,\beta  
,\gamma ,\delta $  all different, to be the same as $G_{ \alpha  
\beta ,\alpha \delta }$.  In  
fact, they are different, and there is indeed  a jump in  
$G_{z_1,z_2}^{x,y}$  at $z_1=1$.)  
  
Nevertheless, the equation one derives for $G_{z_1,z_2}^{x,y}$  
does  
go over into that for $G_{z,1}^{x,y}$  when $z_1=z$, $z_2=1$.  
  
The other reason for us to display Eqs.~47, 48 and 49 here  
explicitly  
is that the ``innermost block approximation" we are just about to apply to  
Eqs.~46$-$56 does not commute with the limit $z\to 1$.  
  
Let us now assume that (45) holds. We can than neglect all the  
${{\cal O}}$($\tau ^2$) terms in (46)-(56)  which means  
dropping all the $uq^2$  
terms, but also the terms in $xqG$  or $q\int G$, except for the  
contributions  
coming from the  
upper end of the replica integrals (the innermost block  
approximation). For example, the penultimate term in Eq.~46  
becomes  
$$2w\int_x^1{{\rm  
d}}tq(t)G_{1,t}^{x,x}\to2wq_1G_{1,x_1}^{x,x}$$  
where we have used the fact that (apart from the jump exactly  
at $z_1,z_2=1$  in the lower indices) whenever an overlap  
variable  
goes beyond the breakpoint $x_1$ it gets stuck to it.  
  
The simplification  gained by dropping the ${{\cal O}}$($\tau  
^2$) terms is  
tremendous: the set of integral equations (46)-(55) reduces to  
the following set of linear algebraic equations:  
\setcounter{equation}{45}  
\primeq  
\begin{eqnarray}  
&&(p^2-2wq_1)G_{1,1}^{x,x}+2wq_1G^{x,x}_{1,x_1}+  
2wq(x)G_{1,x}^{x,x_1}=1\quad ,\\  
%  
% 2  
%  
&&p^2G_{z,1}^{x,x}+wq_1\left (G_{z,x_1}^{x,x}-  
G_{z,1}^{x,x}\right  
)+wq(x)\left (G_{z,x}^{x,x_1}+G_{1,x}^{x,x_1}+G_{x_1,x}^{x,z}-  
G_{1,x}^{x,z}\right )+\nonumber\\  
&&\qquad\qquad +wq(z)\left (G_{1,x_1}^{x,x}-G_{1,1}^{x,x}\right  
)=0,\quad  
1>z\ge x \quad ,\\  
%  
% 3'  
%  
&&p^2G_{1,x}^{x,y}+wq_1\left (G_{x_1,x}^{x,y}-  
G_{1,x}^{x,y}\right  
)+wq(y)\left (G_{y,x}^{x,x_1}+G_{1,x}^{x,x_1}\right  
)+wq(x)\left  
(G_{y,x_1}^{x,x}-G_{y,1}^{x,x}\right. +\nonumber\\  
&&\qquad\qquad +\left . G_{1,x_1}^{x,x}-G_{1,1}^{x,x}\right  
)=0,  
\quad  
x\le y <1\quad ,\\  
%  
% 4'  
%  
&&p^2G_{1,x}^{x,y}+wq_1\left (G_{x_1,x}^{x,y}-  
G_{1,x}^{x,y}\right  
)+wq(x)\left (G_{x,x_1}^{x,y}-G_{x,1}^{x,y}\right )+wq(y)\left  
(G_{y,y}^{x,x_1}+G_{1,x}^{x,x_1}+\right .\nonumber\\  
&&\qquad\qquad +\left .G_{1,x_1}^{x,x}-G_{1,1}^{x,x}\right)=0  
,\quad y \le x < 1\quad ,\\  
%  
% 5'  
%  
&& p^2G_{z,z}^{x,y}+2wq(y)G_{z,z}^{x,x_1}+4wq(z)\left  
( G_{x_1,x}^{x,z}-G_{1,x}^{x,z}\right )=0\quad z \le x,y <1  
\quad ,\\  
%  
% 6'  
%  
&&p^2G_{z,x}^{x,y}+2wq(y)G_{z,x}^{x,x_1}+2wq(z)\left  
(G_{x_1,x}^{x,z}-G_{1,x}^{x,z}\right )+\nonumber\\  
&&\qquad\qquad +2wq(x)\left (G_{z,x_1}^{x,x}-  
G_{z,1}^{x,x}\right  
)=0,  
\quad x\le z \le y <1\quad ,\\  
%  
% 7'  
%  
&&p^2G_{z,z}^{x,y}+wq(y)\left  
(G_{y,y}^{x,x_1}+G_{z,z}^{x,x_1}+2G_{x_1,x}^{x,z}-  
2G_{1,x}^{x,z}\right )+\nonumber\\  
&&\qquad\qquad +2wq(z)\left (G_{x_1,x}^{x,y}-  
G_{1,x}^{x,y}\right  
)=0,  
\quad y\le z \le x <1\quad ,\\  
%  
% 8'  
%  
&&p^2G_{z,x}^{x,y}+wq(y)\left  
(G_{y,x}^{x,x_1}+G_{z,x}^{x,x_1}+G_{x_1,x}^{x,z}-G_{1,x}^{x,z}  
\right )+wq(x)\left (G_{y,x_1}^{x,x}+G_{z,x_1}^{x,x}-\right .  
\nonumber \\ && \qquad\qquad \left  
.-G_{y,1}^{x,x}-G_{z,1}^{x,x}\right  
)+wq(z) \left (G_{x_1,x}^{x,y}-G_{1,x}^{x,y}\right )=0, \quad x  
\le  
y \le z< 1\quad ,\\  
%  
% 9'  
%  
&&p^2G_{z,x}^{x,y}+wq(y)\left  
(G_{y,y}^{x,x_1}+G_{z,x}^{x,x_1}+G_{x_1,x}^{x,z}-  
G_{1,x}^{x,z}+G_{z,x_1}^{x,x}-G_{z,1}^{x,x}\right )+\nonumber\\  
&&\qquad\qquad+w\left (q(x)+q(z)\right )\left (G_{x,x_1}^{x,y}-  
G_{x,1}^{x,y}\right )=0,\quad y\le x \le z <1\quad ,\\  
%  
% 10'  
%  
&& p^2G_{z_1,z_2}^{x,x}+wq(z_1)\left (G_{z_2,x_1}^{x,x}-  
G_{z_2,1}^{x,x}\right )+wq(z_2)\left (G_{z_1,x_1}^{x,x}-  
G_{z_1,1}^{x,x}\right )+\nonumber\\  
&&\qquad\qquad +wq(x)\left  
(G_{z_1,x}^{x,x_1}+G_{x_1,x}^{x,z_1}-G_{1,x}  
^{x,z_1}+G_{z_2,x}^{x,x_1}+\right .\nonumber \\  
&& \qquad\qquad +\left . G_{x_1,x}^{x,z_2}-  
G_{1,x}^{x,z_2}\right  
)=0, \quad  
x \le z_1,z_2 <1\quad .  
\end{eqnarray}  
\normeq\setcounter{equation}{56}  
  
In Eqs.~46'$-$55' we have set $\tau = wq_1$, which is the  
correct stationary condition at this order.  
  
It is evident from Eqs.~46'$-$55' that in the given approximation  
each  
component of $G$ is $\displaystyle\frac{1}{p^2}$ times a  
function  
of $\displaystyle \frac{wq_1}{p^2}$,  $\displaystyle  
\frac{wq(x)}{p^2}$, $\displaystyle \frac{wq(y)}{p^2}$  and  
$\displaystyle \frac{wq(z)}{p^2}$ :  
\begin{eqnarray}  
G=\frac{1}{p^2}g(\frac{wq}{p^2})\quad , \quad p^2 \gg \tau ^2  
\end{eqnarray}  
where $q$ stands  for $q_1$, $q(x)$, $q(y)$, $q(z)$, according  
to the component in question.  
  
The solution of Eqs.~46'$-$55' is particularly simple if all  
overlap variables  
are greater than $x_1$. 
The propagators in this limit will be seen to describe  
fluctuations in a single phase space valley. Then the set (46')$-$  
(55') collapses to 3 simple equations with the solutions:  
\begin{eqnarray}  
G_{1,1}^{x_1,x_1}&=&\frac{1}{p^2}\left  
(1+\frac{2wq_1}{p^2+2wq_1}+\frac{4w^2q_1^2}{(p^2+2wq_1)^2}\right  
)  
\nonumber\\  
G_{1,x_1}^{x_1,x_1}&=&\frac{1}{p^2}\left  
(\frac{1}{2}\frac{2wq_1}{p^2+2wq_1}+\frac{4w^2q_1^2}{(p^2+2wq_1  
)  
^2  
}\right )\\  
G_{x_1,x_1}^{x_1,x_1}&=&\frac{1}{p^2}\frac{4w^2q_1^2}{(p^2+2wq_  
1)  
^2  
}\quad .\nonumber\end{eqnarray}  
  
Comparing these with the replica symmetric propagators given in  
Eq.~35, we see that to the order regarded here (i.e.\ neglecting  
$\tau ^2$ terms ), they are the same. The difference is, of  
course,  
that Eq.~58 is supposed to be valid only for $p^2\gg \tau ^2$, so for the 
time being we keep away from the region where the instability showed  
up.  
  
Two particularly illuminating combinations are the transverse  
and the longitudinal propagators  
\begin{eqnarray}  
\begin{array}{rcl}  
\displaystyle  
G_{\perp}&=&G_{1,1}^{x_1,x_1}-2G_{1,x_1}^{x_1,x_1}+  
G_{x_1,x_1}^{x_1,x_1} =\displaystyle \frac{1}{p^2}\quad ,\\  
G_{\parallel}&=&G_{1,1}^{x_1,x_1}-  
4G_{1,x_1}^{x_1,x_1}+3G_{x_1,x_1}^{x_1,x_1}=  
\displaystyle \frac{1}{p^2+2wq_1}\quad ,  
\end{array}  
\end{eqnarray}  
which are precisely of the form of  the Gaussian propagators in  
a massless phase.  
  
Now we display the complete set of  propagators valid for  
generic overlaps in the near infrared limit $p^2 \gg\tau ^2$:  
\begin{eqnarray}  
G_{1,1}^{x,x}&=&\frac{1}{p^2}\left \{1+\frac{2wq_1}{p^2}+\frac{  
2w^2q_1^2}{p^4}-  
\frac{2w^2q^2(x)}{p^4}\frac{p^4+8wq_1p^2+8w^2q_1^2}{(p^2+2wq_1)^2}+
\right .\nonumber\\
&&+\left .\frac{  
8w  
^4q^4(x)}{p^4(p^2+2wq_1)^2}\right \}\\[2mm]
%  
% 21'  
%  
G_{1,z}^{x,x}&=&\frac{1}{p^2}\left \{\left  
(1+\frac{2wq_1}{p^2}\right )\frac{wq(z)}{p^2}-  
4\frac{wq_1p^2+w^2q_1^2}{(p^2+2wq_1)^2}\frac{w^2q^2(x)}{p^4}-  
\frac{8w^3}{p^4}\frac{q^2(x)q(z)}{p^2+2wq_1}+ \right .\nonumber  
\\  
&& +\left  
.\frac{4w^4}{p^4}\frac{2q^4(x)+q^2(x)q^2(z)}{(p^2+2wq_1)^2} 
\right \}, \quad x \le z <1\\[2mm]
%  
% 22'  
%  
G_{1,x}^{x,y}&=&G_{1,y}^{y,x}=\frac{wq(x)}{p^4}\left \{1+\frac  
{2wq_1}{p^2}-  
4\frac{wq_1p^2+w^2q_1^2}{(p^2+2wq_1)^2}\frac{wq(y)}{p^2}-  
\frac{4w^2}{p^2} \frac{q^2(x)+q^2(y)}{p^2+2wq_1} +\right  
.\nonumber  
\\ && +\left .\frac{4w^3}{p^2}\frac{q^3(y)+2q(y)  
q^2(x)} {(p^2+2wq_1)^2}\right \}, \quad x \le y <1\\[2mm]
G_{1,x}^{x,y}&=&\frac{wq(y)}{p^4}\left \{1+\frac  
{2wq_1}{p^2}-  
4\frac{wq_1p^2+w^2q_1^2}{(p^2+2wq_1)^2}\frac{wq(x)}{p^2}-  
\frac{4w^2}{p^2} \frac{q^2(y)+q^2(x)}{p^2+2wq_1} +\right  
.\nonumber  
\\ &&  +\left .\frac{4w^3}{p^2}\frac{q^3(x)+2q(x)  
q^2(y)} {(p^2+2wq_1)^2}\right \}, \quad y \le x <1\\[2mm]
%  
% 23'  
%  
G_{z,z}^{x,y}&=&  
{\begin{array}[t]{r}  
4\displaystyle\frac{w^2q^2(z)}{p^6}\frac{(p^2+  
2wq_1-2wq(x))(p^2+2wq_1-2wq(y))}{(p^2+2wq_1)^2},\\  
\displaystyle  z \le x,y <1  
\end{array}}\\[2mm]
%  
% 24'  
%  
G_{z,x}^{x,y}&=&  
{\begin{array}[t]{r}  
\displaystyle\frac{4wq(x)}{p^6}(p^2+2wq_1-  
2wq(y))\left (\frac{wq(z)}{p^2+2wq_1}-  
\frac{w^2q^2(x)+w^2q^2(z)}  
{(p^2+2wq_1)^2} \right ),\\  
 x \le z \le y < 1\end{array}}\\[2mm]
%  
% 25'  
%  
G_{z,x}^{x,y}&=&\frac{2wq(x)}{p^6}\left [wq(y)+wq(z)-  
2w^2\frac{q^2(x)+2q(y)q(z)+q^2(y)}{p^2+2wq_1}+\right .\nonumber  
\\  
&&\left  
.+2w^3\frac{q^3(y)+2q^2(x)q(y)+q(y)q^2(z)}{(p^2+2wq_1)^2}\right  
],  
\quad x\le y \le z < 1\\[2mm]
%  
% 26'  
%  
G_{z,x}^{x,y}&=&\frac{2wq(y)}{p^6}\left [wq(x)+wq(z)-  
2w^2\frac{q^2(y)+2q(x)q(z)+q^2(x)}{p^2+2wq_1}+\right .\nonumber  
\\  
&&\left  
.+2w^3\frac{q^3(x)+2q^2(y)q(x)+q(x)q^2(z)}{(p^2+2wq_1)^2}\right  
],  
\quad y\le x \le z < 1\\[2mm]
%  
% 27'  
%  
G_{z,z}^{x,y}&=&G_{z,y}^{y,x}=\frac{4wq(y)}{p^6}(p^2+2wq_1-  
2wq(x))\left (\frac{wq(z)}{p^2+2wq_1}-\right .\nonumber\\  
&&-\left. \frac{w^2q^2(y)+w^2q^2(z)}{(p^2+2wq_1)^2}  
\right ), \quad y\le z \le x< 1\\[2mm]  
%  
% 28'  
%  
G_{z_1,z_2}^{x,x}&=&\frac{2}{p^6}\left  
[w^2q^2(x)+w^2q(z_1)q(z_2)-  
\frac{4w^2q^2(x)(q(z_1)+q(z_2))}{p^2+2wq_1}+\right .\nonumber\\  
&&+\left  
.\frac{2w^4q^2(x)(q^2(z_1)+2q^2(x)+q^2(z_2))}{(p^2+2wq_1)^2}  
\right ], \quad x\le z_1,z_2<1\quad .  
\end{eqnarray}  
  
These formulae can be verified by a direct substitution into  
Eqs.~46'$-$55'.  
  
Note the simple excitation spectrum displayed by Eqs.~58$-$69: the  
propagator  
components have two poles, one at $p^2=-2wq_1$, the other at  
$p^2=0$. When  
we work out the complete expressions for the propagators below,  
we shall see  
that these singularities form, in fact, two continuous bands  
which, having a  
bandwidth of the order of the small mass, cannot, however, be 
resolved in  
the near infrared. In addition to these cuts, we shall also  
find a pole at  
$p^2=0$.  
  
The above formulae, derived under the only assumption of $p^2  
\gg  
\tau ^2$ should also cover the region near the cutoff. Indeed,  
the  
large $p$ expansion of Eqs.~60$-$69 reproduces Eqs.~42, 43 and
44  
provided  
the coefficients of $1/p^2$, $1/p^4$, etc.\ there are evaluated  
in the  
Parisi parametrization to leading order in $\tau$.  
Eqs.~60$-$69 will be used when calculating loop corrections  
for  
$6<d<8$.

\section{The complete solution for the propagators}  
  
The expressions derived in the previous Section are valid in  
the range where the momentum is much larger than the smaller  
mass scale involved in the problem. If we tried to continue  
these formulae into the exceedingly  long wavelength limit,  
i.e.\ to momenta around the small mass scale, $p^2\sim\tau ^2$,  
or  
even  much smaller, $p^2 \ll \tau ^2$, we would encounter truly  
unmanageable infrared singularities, with the propagator  
components  
blowing up like $p^{-6}$. In order to uncover the true  
behaviour in  
this extremely  
long wavelength limit, which we will call the far infrared  
region, we have to return to the original set of
Eqs.~46$-$56 for the propagators. Unfortunately, we have not been  
able  
to devise any approximation scheme (analogous to the ``large $p$  
expansion" or  the ``innermost block approximation" working  
near the upper cutoff and in the near infrared, respectively)  
that would directly give the propagators for the far infrared.  
%(The case of both upper indices of the propagator being zero,  
%i.e.\ of  $G^{0,0}_{z_1,z_2}$, is special, in that it can be  
%obtained in closed form, see below.) For generic overlaps,  
%however,  
Therefore we have to face the task of solving the complete set of  
Eqs.~46$-$56.  
  
This task may, at first sight, appear utterly hopeless.  
Nevertheless, it turned out to be possible to find a solution  
to the  set (46)$-$(56) in closed form  
(even in a more general setting,  with an  
external magnetic field acting on the system) which was  
published  
by two of us in [28], with a minor error corrected in [40].  
  
In retrospect, it is clear that the solvability of the problem  
depends upon the combination of two independent factors. One of  
them is the residual symmetry which allows one to reduce the  
inversion of not only the Hessian, but also of any ultrametric  
matrix to the  inversion of a much simpler object we called the  
kernel in [40,41]. The other is that close to the transition point  
this kernel turns out to be very simple. These two factors  
deserve a separate discussion in their own right which is what  
we present in the two subsections below.  
  
\subsection{The inversion problem of a generic ultrametrix  
matrix}  
  
For the sake of the discussion in this subsection we shall  
assume that both the replica number $n\equiv p_0$  and  the  
block sizes $p_1,p_2,\ldots ,p_R$, featuring in the Parisi  
construction are positive integers such that $p_{i+1}$ is a  
divisor  
of $p_i$, $i=0,1,2,\ldots ,R-1$. This will allow us  
to stay within the limits of well established mathematics  
throughout this subsection. When coming to the application of  
the results  
in the next subsection, we shall have to  
consider, of course, the replica limit $n\to 0$, along with the  
$p_i$'s  
to be continued into the interval $[0,1]$ and $R\to \infty$.  
  
The free energy  $F$ in  the MF approximation is a functional  
of  
the order parameters $q_{\alpha \beta }$, with $q_{\alpha  
\beta }=q_{\beta \alpha }$ and $q_{\alpha \alpha }=0$, $\alpha  
,\beta =1,2,\ldots ,n$. Since  
none of the $n$ replicas is distinguished, $F$ must depend only  
on permutation invariant combinations of the  $q_{\alpha \beta  
}$'s, such as those in Eq.~6, for example. This  means that  $F$ is  
invariant under the action of the elements $\Pi$ of the  
permutation group $S_n$: $F=F(q_{\alpha \beta  
})=F(q_{\scriptscriptstyle \Pi(\alpha  
),\Pi(\beta )})$.  
  
The physical value of the free energy is obtained from the  
functional $F$ by evaluating it at a saddle point where  
$\displaystyle\frac{\partial F}{\partial q_{\alpha ,\beta  
}}=0$.  
The solution of the saddle point equations is a point in the  
$\frac{1}{2}n(n-1)$ dimensional space spanned by the  
independent  
components of $q_{\alpha \beta }$, thus  the order parameter  
$q_{\alpha \beta }$, which is usually spoken of as a  
matrix, is, in fact, a {\it vector}. We do not want to  
introduce a  
separate notation for this vector, but propose to think of it  
as a column vector, whose components are the elements (above the  
diagonal) of $q_{\alpha \beta }$ listed in some fixed order.  
  
The symmetry of $F$ under permutations does not imply that all  
the  solutions of the saddle point equations must  
respect this symmetry. In fact, we know from de  
Almeida and Thouless [31] that at low temperatures the replica  
symmetric saddle point becomes unstable against fluctuations  
that break permutation symmetry. Symmetry breaking gives rise,  
like at any other phase transition, to the reduction of the  
original symmetry group to one of its subgroups. Parisi's  
Ansatz [1] is, in fact, nothing but a concrete choice for this  
residual symmetry, one that, as it turned out later [37],  
embodies the physical assumption of the existence of many  
equilibrium states with an ultrametric organization. For this  
reason we proposed the name ultrametric group for this  
particular residual symmetry in [41].  
  
The vector pointing to the Parisi saddle point is invariant  
under the action of the ultrametric group. The set of  
$\frac{1}{2}n(n-1)\times\frac{1}{2}n(n-1)$  
matrices that, acting on the Parisi $q_{\alpha \beta }$,  
perform  
the  
permutations belonging to the ultrametric group (and thus do  
not change the form of $q_{\alpha \beta } $) constitute an (in  
general, reducible) representation of the group.  
  
Now consider the  $\frac{1}{2}n(n-1)\times\frac{1}{2}n(n-1)$  
dimensional matrices that commute with all  
the matrices belonging to the previous set, i.e.\ with the  
matrices representing the elements of the ultrametric group.  
These are the matrices that we called ultrametric matrices in [41]  
where we also gave a detailed description of their structure.  
If we transform our coordinate system in replica space such as  
to decompose the representation of the group into its  
irreducible components, we find that, by virtue of the Wigner-  
Eckart theorem, in this new representation all ultrametric  
matrices will be block diagonal.  
  
The Hessian or stability matrix (essentially the inverse  
propagator) is an important example of an ultrametric matrix.  
Constructing the irreducible representations of the residual  
symmetry group one can therefore make  significant progress  
towards the diagonalization and inversion of the Hessian or of  
any  
other ultrametric matrix.  
  
The approach indicated above is {\it the} standard one. An  
alternative that may involve seemingly ad  hoc steps but leads  
to perfectly identical results is to directly construct the  
orthogonal subspaces which are closed under the action of any  
ultrametric matrix and obtain the block diagonal form by  
transforming the Hessian to this new basis. This is what we did  
in [42], exploiting the consequences of this block  
diagonalization fully in [41]. (The decomposition of
the reducible representation mentioned above into irreducible
parts  
has since been constructed by B\'antay and  
Zala [43].)  
  
A most detailed account of our approach can be found in [41].  
The same results were somewhat later reproduced via purely  
algebraic means by De Dominicis, Carlucci and Temesv\'ari [44]  
by using  an efficient tool called the replica Fourier  
transform (RFT).  
  
Whatever the procedure, the central result one obtains is the  
following. In the new basis the Hessian breaks up into a string  
of $(R+1)\times (R+1)$  blocks followed by a string of $1\times  
1$  
``blocks" along the diagonal. The $(R+1)\times (R+1)$ blocks are  
called, for reasons of little interest, longitudinal-anomalous  
(LA), and they are labelled by an index $k=0,1,2,\ldots ,R+1$.  
The  
sector where the $1\times 1$ blocks appear, i.e.\ where the  
transformation actually diagonalizes the Hessian, is called the  
replicon (R) sector.  
  
Let us call our generic ultrametric matrix (in the original  
coordinate system) $\underline{\underline {{\bf M}}}$ and its  
inverse $\underline{\underline {{\bf G}}}$. Their components  
can  
be parametrized as in Eq.~23. Let us further call the  
corresponding matrices in the new representation  
$\underline{\underline {\hat {{\bf M}}}}$ and  
$\underline{\underline {\hat {{\bf G}}}}$,  
respectively. As shown in [41,42], the  
diagonal elements of $\underline{\underline {\hat {{\bf  
M}}}}$ in  the replicon sector are given by:  
\begin{eqnarray}  
%{}_R
{{\hat {{M}}}}^{r,r}_{k,l}=  
\sum_{s=k}^{R+1}p_s\sum_{t=l}^{R+1}p_t(M^{r,r}_{t,s}-M^{r,r}_{t-  
1,s}  
-M^{r,r}_{t,s-1}+M^{r,r}_{t-1,s-1})\quad ,  
\end{eqnarray}  
where the discrete indices $r$, $k$, $l$ needed to label these  
matrix elements take on the values  $r=0,1,\ldots ,R$, and  
$k,l=  
r+1,r+2,\ldots ,R+1$, respectively. Similar formulae hold, of  
course, for $
%{}_R
\underline{\underline {\hat {{\bf G}}}}$:  
\begin{eqnarray}  
%{}_R
{{\hat {{G}}}}^{r,r}_{k,l}=  
\sum_{s=k}^{R+1}p_s\sum_{t=l}^{R+1}p_t(G_{t,s}^{r,r}-G_{t-  
1,s}^{r,r}-  
G_{t,s-1}^{r,r}+G_{t-1,s-1}^{r,r})\quad .  
\end{eqnarray}  
  
The combinations appearing in Eqs.~70 and 71 can alternatively be  
regarded as (double) RFT's [44]; in that context, to  
conform to earlier notation, it is useful to denote them as  
\begin{eqnarray}  
\begin{array}{ll}  
\displaystyle 
%{}_R
\hat M^{r,r}_{k,l}\equiv K^{r,r}_{k,l}\quad ,  
&  
\\ & \\  
\displaystyle 
%{}_R
\hat G^{r,r}_{k,l}\equiv F_{k,l}^{r,r} 
\quad ,& r+1\le k,l\le R+1.  
\end{array}  
\end{eqnarray}  
  
The idea of a transform sharing the convolution property with ordinary 
Fourier transforms was first proposed, for the continuum limit, by M\'ezard 
and Parisi [45] in the context of random manifolds. 
It was later recognized by Parisi and Sourlas [46] as a Fourier transform 
within the p-adic number approach (again limited to $R\to\infty $, $n\to 
0$). In this context Parisi and Sourlas could derive, for the replicon 
sector, the relationship $F_{k,l}^{r,r}\cdot K_{k,l}^{r,r}=1$ which in our 
geometric (or group theoretic) approach follows from the fact that  
$\matr {M}\cdot\matr {G}=\matr {1}$ and that these matrices are diagonal in 
that sector. The extension of the RFT to the discrete case and to the LA 
sector through the use of generalized Parisi boxes $p_t^{(r,s)}$, as described 
below, was made in [44]. Finally, the connection to standard Fourier 
transforms over a group was given in [47].

Turning to the longitudinal-anomalous sector, there we have the  
$(R+1)\times(R+1)$ diagonal blocks labelled by the index  
$k=0,1,\ldots  
,R+1$.  
The matrix elements work out to be
\footnote{In order to make contact with earlier notation,
we note that the quantity
$\hat M^{r,r}_{k,l}$
we use here was called $\lambda (r;k,l)$ in [41]; ${{\hat
{{M}}}}^{r,s}_{k}$ here is related to $M^{(k)}_{r,s}$ in [41]
by a similarity transformation, and the relationship between
the quantity $\Delta_r^{(k)}$ in [41] and $\delta_r^{(k-1)}$
here is $\Delta_r^{(k)}=\frac{1}{2}\delta_r^{(k-1)}$. The kernel
is, however, invariant, so it is exactly the same as the quantity
called $K_k(r,s)$ in [41], and the Dyson equation (81) also
remains the same.}
\begin{eqnarray}  
%{}_{LA}
{{\hat {{M}}}}^{r,s}_{k}=  
\Lambda _k(r)\delta ^{Kr}_{r,s}+\frac{1}{4}K^{r,s}_k\delta  
_s^{(k-  
1)}\quad ,&& r,s=0,1,\ldots ,R\quad ,  
\end{eqnarray}  
where $\Lambda_k(r)$ is  shorthand for  
%\begin{eqnarray}  
\be
\Lambda_k(r)=\left\{
\begin{array}{ll}\displaystyle
%{}_R
{\hat  M}^{r,r}_{k,r+1}\quad ,&\qquad\qquad k>r+1\\
{\hat  M}^{r,r}_{r+1,r+1}\quad ,&\qquad\qquad k\leq r+1
\end{array}\right.
\ee
\be
\delta ^{(l)}_s=p_s^{(l)}-p_{s+1}^{(l)}\quad ,\qquad  
l = 0,1,\ldots , R+1\quad ,\quad s = 0,1,\ldots , R
\ee
and  
\be  
p_s^{(l)}=\left \{  
\begin{array}{ll}  
\displaystyle p_s\quad ,&\qquad\qquad s\leq l\\  
2p_s\quad ,&\qquad\qquad s>l  
\end{array}\right. \quad .  
\ee
  
We shall refer to the objects $K$ and $F$ as the kernel or RFT  
of  
$\matr {M}$ and $\matr {G}$, respectively.  
  
$K_k^{r,s}$ is given in terms of the original matrix elements as  
\begin{eqnarray}  
K_k^{r,s}=\sum_{t=k}^{R+1}p_t^{(r,s)}(M_t^{r,s}-M_{t-  
1}^{r,s})\quad .  
\end{eqnarray}  
  
It is here that we make use of the observation made below Eq.~23,  
and  
keep only the larger of the two lower indices of $\matr M$.  
Formula  
(77)  
could not have been cast into the simple form above had we kept  
to the original, redundant parametrization.  
  
The weight $p_t^{(r,s)}$ is defined as  
\begin{eqnarray}  
p_t^{(r,s)}=\left \{\begin{array}{ll}  
p_t,&t\leq r \leq s\\  
2p_t& r<t\leq s\\  
4p_t& r \leq s < t  
\end{array}\right. \quad . \end{eqnarray}  
  
The corresponding formulas for $
%{}_{LA}
\hat G_k^{r,s}$ are  
\begin{eqnarray}  
%{}_{LA}
\hat G_k^{r,s}=\frac{1}{\Lambda _k(r)}\delta  
_{r,s}^{Kr}+  
\frac{1}{4}F_k^{r,s}\delta _s^{(k-1)},  
\end{eqnarray}  
with the kernel of $\matr{G}$ given by  
\begin{eqnarray}  
F_k^{r,s}=\sum_{t=k}^{R+1}p_t^{(r,s)}(G_t^{r,s}-G_{t-  
1}^{r,s})\quad .  
\end{eqnarray}  
  
Now the equation ${\underline {\underline {{\bf G}}}}  
\cdot{\underline{\underline {{\bf M}}}=\underline {\underline  
{{\bf 1}}}}$  expressed  
in terms of the diagonal blocks reads  
$$  
\sum_{t=0}^R(\delta _{r,t}^{Kr}\frac{1}{\Lambda_k  
(t)}+\frac{1}{4}F_k^{r,t}\delta _t^{(k-1)})(\delta  
_{t,s}^{Kr}\Lambda_k(t)+\frac{1}{4}K_k^{t,s}\delta _s^{(k-  
1)})=\delta _{r,s}^{Kr}\quad ,$$ or  
\begin{eqnarray}  
F_k^{r,s}=-  
\frac{1}{\Lambda_k(r)}K_k^{r,s}\frac{1}{\Lambda_k(s)}-  
\frac{1}{\Lambda_k(s)}\sum_{t=0}^RF_k^{r,t}\frac{\delta _t^{(k-  
1)}}{4}K_k^{t,s}\quad .  
\end{eqnarray}

If we regard $\underline{\underline {{\bf M}}}$ as the  
self-energy matrix and  $\underline{\underline {{\bf G}}}$ the  
propagator then Eq.~81 is just the Dyson equation connecting  
their respective kernels or replica Fourier transforms. This  
Dyson equation was first obtained in [40] for the continuum  
limit and in [41] for the discrete case.  
  
Given the matrix $\underline{\underline {{\bf M}}}$ to be  
inverted,  
 $K_k^{r,s}$ can be computed  
using Eq.~77. Eq.~81 is then a set of matrix equations in the  
unknown  
$ F_k^{t,s}$ (one for each value of $k$). Suppose we  
are able to solve it. Next we have to  
invert the relation (80) to get $\matr{G}$ itself in the LA  
sector, and  
invert Eq.~71 to get it in the R sector. One of the great  
merits of the  
RFT approach is that these inversion formulae  
work out to be fairly transparent. The inverse of Eq.~80 is  
\begin{eqnarray}  
{}_AG_t^{r,s}= \sum_{k=0}^t\frac{1}{p_k^{(r,s)}}(F_k^{r,s}-  
F_{k+1}^{r,s})\quad ,  
\end{eqnarray}  
that of the double transform in Eq.~71 is  
\begin{eqnarray}  
{}_RG_{u,v}^{r,r}=\sum_{k=r+1}^u\frac{1}{p_k}\sum_{l=r+1}^v  
\frac{1}{p_l}(F_{k,l}^{r,r}-F_{k+1,l}^{r,r}-  
F_{k,l+1}^{r,r}+F_{k+1,l+1}^{r,r})\quad .  
\end{eqnarray}  
As for  ${}_{A}G_{u,v}^{r,r}$ we find
\setcounter{equation}{82}\primeq
\be
{}_{A}G_{u,v}^{r,r}={}_AG_u^{r,r}+{}_AG_v^{r,r}
-{}_AG_r^{r,r} \quad ,
\ee
whence the full propagator is obtained as
\normeq
\bea
\begin{array}{rcl}
G^{r,s}_t&=&{}_AG_t^{r,s}\quad,\\
\\
G^{r,r}_{u,v}&=&{}_RG^{r,r}_{u,v}+{}_AG^{r,r}_{u,v}\quad .
\end{array}
\eea

Now we proceed to apply the above formalism to the special case  
of  
inverting (17) in order to determine the free propagator.  
  
%The full propagator is obtained as the sum of the LA and R  
%contributions:  
%\begin{eqnarray}  
%G=G_{LA}+G_R.  
%\end{eqnarray}  
  
\subsection{The exact solutions for the Gaussian propagators  
near  
the transition temperature}  
  
Let us now take the inverse bare propagator for the ultrametric  
matrix  $\matr {M}$  of the previous subsection. As seen from  
Eq.~17, its components are  
\begin{eqnarray}  
\begin{array}{rcl}  
\displaystyle p^2+
%{}^0
M^{r,r}_{R+1,R+1}&=&p^2-2\tau -2uq_r^2,\\  
&&\\  
%{}^0
M^{r,s}_{R+1}&=&-wq(\min\{r,s\}),\\
&&\\
M^{r,r}_{R+1,t}&=&-wq_t,
\end{array}& &  
\begin{array}{rcl}  
\displaystyle r&=&0,1,\ldots ,R\\  
&&\\  
r,s&=&0,1,\ldots,R\\
&&\\
r<t&=&1,2,\ldots,R\quad .  
\end{array}  
\end{eqnarray}  
In the truncated model (2) all other components are zero.  
  
Now we substitute these into Eq.~70 and take the continuum  
limit, i.e.\ perform the analytic continuation in  the $p$'s  
into  
the interval (0,1) and let  $R\to \infty$, according to  
Parisi's  
prescription [1]. In the new representation we get for the  
diagonal components in the replicon sector  
\begin{eqnarray}  
%{}^0_R
\hat M^{x,x}_{t,s}&  
\equiv & K_{t,s}^{x,x}=\left (F_{t,s}^{x,x}\right )^{-1}  
=p^2+\lambda ^{(0)}(x;s,t)\equiv\nonumber  
\\&\equiv & p^2+uq^2(s)+uq^2(t)-2uq^2(x),\qquad 0\le x<s,t\le 1,  
\end{eqnarray}  
where we have also used the equation of state (26). Since in the  
replicon  sector $\hat \matr{M}$ is diagonal, the $\lambda  
^{(0)}$ appearing in Eq.~86  
are nothing but the eigenvalues of the Hessian matrix  
$\matr{M}^{(0)}$  
defined in Eq.~17. These eigenvalues fill a continuous band with  
lower band edge $\lambda ^{(0)}(x;x,x)=0$ and upper band edge  
$\lambda ^{(0)}(0;1,1)=2uq_1^2=\displaystyle \frac{2u}{w^2}\tau  
^2+\cdots$ \begin{eqnarray}  
0\le \lambda ^{(0)}(x;s,t)\le\frac{2u}{w^2}\tau ^2.  
\end{eqnarray}  
  
The LA kernel corresponding to the simple ultrametric matrix  
(85) is obtained from Eq.~77 as 
\begin{eqnarray}  
%{}^0
K_k^{s,t}=-4wq(\min(\{s,t\}),&&  
0\le s,t \le 1 \quad .
\end{eqnarray}  
Note that this is independent of the lower variable  $k$. (If  
we had also kept the Tr($\phi ^4$) invariant in Eq.~2, as in [40], we  
would have a $k$-dependent kernel.)  
  
In the continuum limit Eq.~81 then becomes  
\begin{eqnarray}  
F_k^{x,y}=\frac{4wq(\min\{x,y\})}{\Lambda _k^{(0)}(x)\Lambda  
_k^{(0)}(y)}-\frac{2w}{\Lambda  
_k^{(0)}(y)}\int_0^1\Delta_t^{(k)}q(\min\{y,t\})F_k^{x,t},&&\\  
0\le k,x,y\le 1,\nonumber  
\end{eqnarray}  
where  
\begin{eqnarray}  
\Delta_t^{(k)}=\left (\frac{1}{2}\theta (k-t)+\frac{k}{2}\delta  
(k-t)+\theta (t-k)\right ){{\rm d}}t  
\end{eqnarray}  
is the continuum limit of 
%(75) and  
$-\frac{1}{2}\delta^{(k-1)}_t$
and \footnote{To avoid confusion we note that Eq.~90 is also
the continuum limit of the {\it negative} of a quantity
which was, unfortunately, denoted by the same symbol
$\Delta^{(k)}_t$ in [41].}
\begin{eqnarray}  
\Lambda _k^{(0)}(x)=\left \{  
\begin{array}{lcl}  
\displaystyle p^2+\lambda ^{(0)}(x;x,k)&,&x\le k\\  
\displaystyle p^2&,&x>k.  
\end{array}\right.  
\end{eqnarray}  
  
The kernel in Eq.~88 has a remarkable property, namely  
that it depends only on one (here, the smaller) of its upper  
variables. This is, in fact, the other crucial factor (in  
addition to the symmetry analysed in the previous subsection)  
that allows one to obtain the bare propagator in closed form.  
Indeed, for a kernel depending only on the smaller of its upper  
variables the solution to Eq.~89 becomes a product of two  
factors, one depending on $x$, the other on $y$.  
  
We write it in the following form  
\begin{eqnarray}  
F_k^{x,y}=\frac{4}{W_k}\frac{\Phi _k^{+}(\min\{x,y\})}{\Lambda  
_k^{(0)}(\min\{x,y\})}\frac{\Phi _k^{-}(\max\{  x,y\})}{\Lambda  
_k^{(0)}(\max\{  x,y\})}  
\end{eqnarray}  
where $\Phi _k^{\pm}$ are two independent solutions of  
\begin{eqnarray}  
\frac{{{\rm d}}}{{{\rm d}}x}\frac{1}{w\dot q(x)}\frac{{{\rm  
d}}}{{{\rm d}}x}\Phi _k^{\pm}(x)=-\frac{\Phi  
_k^{\pm}(x)}{2\Lambda ^{(0)} _k(x)}  
\end{eqnarray}  
and the Wronskian  
\begin{eqnarray}  
W_k=\frac{\dot\Phi _k^{+}(x)\Phi _k^{-}(x)-\Phi  
_k^{+}(x)\dot\Phi  
_k^{-}(x)}{w\dot q(x)}  
\end{eqnarray}  
is independent of $x$. (The dot means derivative w.r.t.\ the  
argument).  
  
For the details of the derivation of $\Phi _k^{\pm}$ we refer  
the  
reader to [28,40], here we content ourselves with simply giving  
the results for $F_k^{x,y}$. In order to simplify the  
notation slightly we introduce  
\begin{eqnarray}  
\hat p^2\equiv \frac{u}{w^2}p^2=\frac{x_1}{2wq_1}p^2.  
\end{eqnarray}  
  
With Eqs.~86 and 91 $\Lambda _k^{(0)}(x)$ now becomes  
\begin{eqnarray}  
\Lambda _k^{(0)}(x)=\left \{  
\begin{array}{lcl}  
\displaystyle p^2\left (1+\frac{k^2-x^2}{4\hat p^2}\right  
)&,&x\le k\\  
\displaystyle p^2&,&x>k.  
\end{array}\right .  
\end{eqnarray}  
  
$F_k^{x,y}$ will be expressed in terms of the solutions  
${{{\cal  
G}}}_1$ and ${{{\cal G}}}_2$ of the Gegenbauer equation  
\begin{eqnarray}  
(1-\xi ^2)\ddot {{{\cal G}}}=2{{{\cal G}}}\end{eqnarray}  
belonging to the initial conditions  
\begin{eqnarray}  
\begin{array}{lclcl}  
\displaystyle {{{\cal G}}}_1\equiv C(\xi )&,&C(0)=1&,&\dot  
C(0)=0\\  
\displaystyle {{{\cal G}}}_2\equiv S(\xi )&,&S(0)=0&,&\dot  
S(0)=1.  
\end{array}  
\end{eqnarray}  
  
The variable $\xi $ is related to the original
%replica variable
overlap variable
$x$ by  
\begin{eqnarray}  
\xi =\frac{x}{2\hat p}\left (1+\frac{k^2}{4\hat p^2}\right )^  
{-\frac{1}{2}},  
\end{eqnarray}  
and we will frequently use abbreviations like  
$C(\xi )\equiv C_x$ or $C_k\equiv C_{x=k}$ etc.~.  
  
We now have  
\begin{eqnarray}  
p^2F_k^{x,y}&=&\frac{2}{\hat p}\cdot\frac{1}{1+\displaystyle  
\frac{k^2-x^2}{4\hat p^2}}\cdot\frac{1}{1+\displaystyle  
\frac{k^2-  
y^2}{4\hat p^2}}\cdot\frac{A_k}{A_k\sigma _k+B_k\gamma _k}  
\cdot\frac{S_x}{S_k}\cdot\nonumber\\  
&&\!\!\!\!\!\!\!\!\!\!\!\!\!\!\!\!\!\!\!\!\!\!\!\!  
\!\!\!\!\!\!\!\!\!\!\!\!\!\!\!\!\!\!\!\!\!\!\!\!  
\cdot \left [(A_k\sigma _k+B_k\gamma _k)C_y-  
\left(2\sqrt{1+\displaystyle \frac{k^2}{4\hat p^2}}\left  
(\sigma  
_k+\frac{k}{2\hat p}\gamma _k\right )C_k+\gamma _k\dot  
C_k\right  
)S_y\right ],\quad \!\!\!\!\!\!\!  
x\le y\le k \le x_1,\\  
p^2F_k^{x,y}&=&\frac{2}{\hat p}\cdot\frac{1}{1+\displaystyle  
\frac{k^2-x^2}{4\hat p^2}}\cdot\frac{A_k}{A_k\sigma  
_k+B_k\gamma  
_k}  
\frac{S_x}{S_k}\cdot\gamma _y\quad ,\qquad\qquad\!\!\!
%\hskip 8em minus8em
x\le k\le y \le  
x_1,\\  
p^2F_k^{x,y}&=&\frac{2}{\hat p}\cdot\frac{A_k\cosh\left  
(\displaystyle \frac{x-k}{\hat p}\right )+B_k\sinh\left  
(\displaystyle \frac{x-k}{\hat  
p}\right )}{A_k\sigma _k+B_k\gamma _k}\cdot\gamma _y\quad  
,\qquad  
\!\!\!\!\!\!\! k\le x\le y \le x_1,  
\end{eqnarray}  
where the notations  
\begin{eqnarray}  
A_k&=&2\sqrt{1+\displaystyle \frac{k^2}{4\hat p ^2}} \!\!\!\!  
\quad S_k \quad ,\\  
B_k&=&\dot S_k+\frac{k}{\hat p}\sqrt{1+\displaystyle  
\frac{k^2}{4\hat p^2}}\!\!\!\! \quad  
S_k\quad,\\  
\gamma _x&=&\cosh\left (\frac{x_1-x}{\hat p}\right )+\frac{1-  
x_1}{\hat p}\sinh \left (\frac{x_1-x}{\hat p}\right ),\\  
\sigma _x&=&\sinh\left (\frac{x_1-x}{\hat p}\right )+\frac{1-  
x_1}{\hat p}\cosh \left (\frac{x_1-x}{\hat p}\right )  
\end{eqnarray}  
have been introduced.  
  
$F_k^{x,y}$ is a continuous function of $x$, $y$ and $k$.  For  
$y<x$, Eqs.~100$-$102 apply with $x$ and $y$ interchanged. Whenever  
either $x$, $y$ or $k$ goes beyond $x_1$, $F_k^{x,y}$ becomes a  
constant in that variable.  
  
In the above notation the Wronskian in Eq.~94 works out to be  
\begin{eqnarray}  
W_k=A_k\sigma _k+B_k\gamma _k \quad .
\end{eqnarray}  
  
It can be shown that the roots of  
\begin{eqnarray}  
W_k(p^2=-\lambda )&=&0  
\end{eqnarray}  
give the LA eigenvalues of the Hessian $\matr {M}^{(0)}$.  
  
Writing up Eq.~107 explicitly with the help of the definitions  
(103)-(106) we can show that  Eq.~108 has infinitely many  
positive roots for any given $0\le k \le x_1$. These can be  
labelled by a discrete index as $\lambda _m(k)$,  
$m=0,1,2\ldots$.  
The eigenvalues belonging to $m=0$ are of the order $\tau $  
and,  
with $k$ varying between 0 and $x_1$, they form a continuous  
band:  
\begin{eqnarray}  
2\tau \left (1-\frac{u}{3w^2}\tau +\cdots \right )<\lambda  
_0(k)  
<2\tau \left (1+\frac{u}{3w^2}\tau +\cdots \right ).  
\end{eqnarray}  
All  the other eigenvalues are of ${{\cal O}}(\tau ^2)$. For a given $k$  
they fall off like $\sim\frac{\tau ^2}{m^2}$, while for a given  
$m$ they form a continuous band of  
width of ${{\cal O}}(\tau ^2)$ as $k$ is varied. The bands corresponding  
to $m=1,2,\ldots$ partially overlap and even the largest  
($\lambda _1(x_1)$) of them is smaller than the upper edge of  
the  
replicon band given in Eq.~87.  
Therefore the small,  ${{\cal O}}(\tau ^2)$ eigenvalues can be regarded  
as  
forming a single continuous band spanning the range  
[$0,\frac{2u}{w^2}\tau ^2$].  
  
Replica symmetry breaking resolves part of the degeneracy of  
the  
AT eigenvalues: the large eigenvalue is split and smeared out  
into the narrow band (109), while the small (negative)  
eigenvalue, in addition to being split, also gets shifted, so  
that the smallest eigenvalue  is now zero: RSB has cured the AT  
instability.

It is clear that these continuous bands of eigenvalues will  
show up as narrow branch cuts in the propagators. In the near  
infrared limit, with squared momenta much larger than the  
bandwidths, these cuts appeared  as simple poles. We have now  
learnt the precise structure of the singularities so we are  
prepared to go into the far infrared limit. Before turning to  
this we write up the various propagator components with the help  
of the (continuous forms) of the inversion formulas given in  
Eqs.~82, 83 and 83'.  
  
The various components of the propagator will be displayed with  
the LA and R contributions added up.  
  
For the ``first propagator" (two pairs of replica indices  
coinciding) we have:  
\begin{eqnarray}  
G_{1,1}^{x,x}=& -& \left (\int _0^x+\frac{1}{2}\int  
_x^{x_1}\right  
) \frac{{{\rm d}}k}{k}\frac{\partial }{\partial  
k}F_k^{x,x}+\frac{1}{2}F_{x_1}^{x,x}+\int _x^{x_1}\frac{{{\rm  
d}}k}{k}\frac{\partial }{\partial k}\int_x^{x_1}\frac{{{\rm  
d}}l}{l}\frac{\partial }{\partial l}F_{k,l}^{x,x}-\nonumber \\  
& -&  \int_x^{x_1}\left ( \frac{{{\rm d}}k}{k}\frac{\partial}  
{\partial k}F_{k,x_1}^{x,x}+\frac{{{\rm d}}l}{l}\frac{\partial  
}{\partial l}F_{x_1,l}^{x,x}\right )+F_{x_1,x_1}^{x,x}\quad .  
\end{eqnarray}  
  
For the ``second propagator" (one pair of replica indices  
coinciding) we have two different formulae according to the  
relative values of the variables:  
\begin{eqnarray}  
G_{1,x}^{x,y}=&-&\left (\int  
_0^x+\frac{1}{2}\int_x^y+\frac{1}{4}\int _y^{x_1}\right  
)\frac{{{\rm d}}k}{k}\frac{\partial }{\partial  
k}F_k^{x,y}+\frac{1}{4}F_{x_1}^{x,y}\quad , \qquad x \le y <  
1,\\  
G_{1,t}^{x,x}=&-&\left  
(\int_0^x+\frac{1}{2}\int_x^t+\frac{1}{4}\int _t^{x_1}\right  
)\frac{{{\rm d}}k}{k}\frac{\partial }{\partial  
k}F_k^{x,x}+\frac{1}{4}F_{x_1}^{x,x}+\int_x^{x_1}\frac{{{\rm  
d}}k}{k}\frac{\partial }{\partial k}\int _x^t\frac{{{\rm  
d}}l}{l}\frac{\partial }{\partial l}F_{k,l}^{x,x}-\nonumber \\  
&-&\int_x^t\frac{{{\rm d}}k}{k}\frac{\partial }{\partial  
k}F_{k,x_1}^{x,x}\quad ,\qquad x \le t < 1.  
\end{eqnarray}  
  
Finally, for the ``third propagator" (all replica indices  
different) we find four different expressions, depending again  
on the  order of the variables:  
\begin{eqnarray}
G^{x,y}_t=  
G_{x,t}^{x,y}
=G^{y,x}_{y,t}
&=&-\left (\int  
_0^x+\frac{1}{2}\int_x^y+\frac{1}{4}\int _y^{t}\right  
)\frac{{{\rm d}}k}{k}\frac{\partial }{\partial k}F_k^{x,y}\quad  
,  
\; x \le y \le t < 1,\\
G^{x,y}_t=
G_{x,t}^{x,y}
=G^{y,x}_{t,t}
&=&-\left (\int _0^x+\frac{1}{2}\int_x^t\right  
)\frac{{{\rm d}}k}{k}\frac{\partial }{\partial k}F_k^{x,y}\quad  
,  
\,\:\quad\qquad\, x \le t \le y < 1,\\
G^{x,y}_t=
G_{t,t}^{x,y}
=G^{y,x}_{t,t}
&=&-\int _0^t \frac{{{\rm d}}k}{k}\frac{\partial  
}{\partial k}F_k^{x,y}\quad ,
\qquad\qquad\qquad\qquad\, t \le x  \le y < 1,\\  
G_{s,t}^{x,x}&=&-\left (\int  
_0^x+\frac{1}{2}\int_x^s+\frac{1}{4}\int _s^{t}\right  
)\frac{{{\rm d}}k}{k}\frac{\partial }{\partial k}F_k^{x,x}+  
\nonumber \\  
&&+\int_x^s\frac{{{\rm d}}k}{k}\frac{\partial }{\partial k}  
\int  
_x^t\frac{{{\rm d}}l}{l}\frac{\partial }{\partial  
l}F_{k,l}^{x,x}  
\quad , \qquad x \le s,t < 1.  
\end{eqnarray}  
  
In Eqs.~110$-$116 $F_k^{x,y}$ is the LA kernel given by  
Eqs.~100$-$102 and $F_{k,l}^{x,x}$ is the R kernel defined in 
Eq.~86.  
  
The various propagator components are continuous  functions of  
all overlap variables, except when either or both lower indices  
take on the value 1 corresponding to coinciding replica  
indices.

\section{The far infrared region}  
  
The formulae derived in the previous section give a complete  
solution for the propagators of the  
Ising spin glass near the transition temperature
and in zero field. Expanding  
them in the limit  $p^2\gg uq_1^2$ one can 
easily recover the results, derived by different means in Sec.~6, 
for the {\it near infrared} region. To extract  
information concerning the {\it far infrared} ($p^2 \sim uq_1^2$ or  
$p^2 \ll u q_1^2$) is  
considerably  
harder. The main difficulty consists in  
the extreme richness of behaviour shown by the various  
propagator components. Indeed, depending  
on the relative magnitude of the overlap variables we have such  
a large number of different analytic  
forms that any attempt to display an exhaustive set of results  
would be quite illusory. What we are  
able to do is to merely give a sample of the results which will  
illustrate the types of behaviours one  
encounters in the extreme long wavelength limit.  
  
Let us start with the simplest propagators, with all overlap
variables above the breakpoint (the  
``single-valley propagators"). With some effort one finds from  
the complete formulae in Sec.~7 that for  
$p^2 \ll uq_1^2$, i.e.\ far below the upper edge of the small 
 mass band, these components are given by the  
following simple expressions:  
\begin{eqnarray}  
G_{1,1}^{x_1,x_1}&\approx&\frac{3}{p^2}\quad ,\\  
G_{1,x_1}^{x_1,x_1}&\approx&\frac{3}{2p^2}\quad ,\\  
G_{x_1,x_1}^{x_1,x_1}&\approx&\frac{1}{p^2}\quad .  
\end{eqnarray}  
The remarkable fact about Eqs.~117-119 is that they are the same  
as the  $p^2\to 0$  limits of the formulae in  
Eq.~58 which were derived in the {\it opposite} limit, $p^2 \gg 
uq_1^2$.  Although there are some (explicitly  
known but complicated) corrections around $p^2\sim uq_1^2$, 
the order of magnitude of  
the propagators is still $\sim 1/p^2$, which  
means that the formulae given in Eq.~58 can be regarded as a good  
representation of the single-valley  
propagators in the entire momentum range. This was first  
pointed out in [30],
but the $p^{-2}$ like IR  
behaviour of the  
single valley propagators had been  
know already from [27].  
  
We are not aware of any {\it a priori} reason why any propagator  
component should behave the same way both in the near and in the  
far infrared (and, indeed, no such continuity is observed in the  
other components), so we believe this coincidence carries a  
physical message. We shall return to this  
point in the next section.  
  
From the continuity of the single-valley propagators one would  
tend to infer the same for the  
transverse and the longitudinal combinations given in Eq.~59.  
This is certainly true for the transverse  
component since, as shown already in [28],
$G_{\perp}={1/p^2}$ is an exact  
result within the Gaussian approximation. As for the longitudinal 
component, its behaviour in the far  
infrared is less clear. The point is that the leading terms  
cancel in the particular combination $G_{\parallel}$ and  
next to leading terms are not controlled  reliably by the  
truncated model that we are using here.   
Nevertheless, it is safe to say that the longitudinal  
propagator is less singular than $1/p^2$ for 
$p\to 0$.
 
It is interesting to note that
in some important respects (masslessness, the ratios
3:$\frac{3}{2}$:1 between the three propagator components, 
weaker singularity in $G_{\parallel}$) the qualitative  
behaviour of the single-valley propagators of Eqs.~117$-$119
is rather similar to that  
of the correlation functions predicted by the droplet theory  
[4,5], despite the obvious conflict between
the underlying physical pictures.  
  
Let us now turn to the other propagator components. For the  
diagonal propagator $G_{1,1}^{x,x}$ we find:  
\begin{eqnarray}  
G_{1,1}^{x,x}\approx \frac{u}{w^2}\left (\frac{1}{x\hat p^3}-  
\frac{1}{2x^2\hat p^2}\right ) , \qquad 0<x<x_1, \quad p\to  
0\quad,  
\end{eqnarray}  
where $\hat p^2$ is the rescaled momentum introduced in Eq.~95.  
Although $G_{1,1}^{x,x}$ is evidently continuous at $x=x_1$,   
Eq.~120 does not match Eq.~117. The apparent contradiction is  
resolved by noting that the limit $p\to 0$ is  
nonuniform: near $x=x_1^{-}$, $G^{x,x}_{1,1}$  depends on combinations like  
$(x_1-x)/p$ and $(x_1-x)^{1/2}/p$. The limit $p\to 0$ is nonuniform also  
around $x=0$, so for a fixed value of $x\in (0,1)$ Eq.~120 will  
hold only if the wavenumber is sufficiently small to satisfy  
both $p\ll x$ and $p \ll (x_1-x)$. Similar remarks apply in 
all the cases that follow.  
  
For the propagator component $G_{1,z}^{x,x}$, $x<z$, we find  
\begin{eqnarray}  
G_{1,z}^{x,x}=\frac{u}{w^2}\left (\frac{1}{x\hat p^3}-  
\frac{1}{2x^2\hat p^2}\right )\quad \mbox{{\rm if}} \quad  
0<x<z\le x_1\quad, \quad p\to 0.  
\end{eqnarray}  
Again, in Eq.~121  $(z-x)$ is understood to be large compared with $p$.  
  
The component $G_{1,x}^{x,y}$, $0<x\le y \le x_1$ tends to a  
constant  
\begin{eqnarray}  
G_{1,x}^{x,y}\to -\frac{4u}{w^2}\frac{1}{x_1(x_1^2-  
x^2)}\frac{S(x/x_1)}{S(1)}\quad \mbox{{\rm if }} 0<x<y=x_1\quad  
,\quad p\to 0\quad . \end{eqnarray}  
The divergence of this constant for $x\to x_1-0 $ is a signal of  
the IR divergence of the limit $x=y=x_1$,  as given by  
Eq.~118. For $x<y<x_1$ and $p\to 0$, $G_{1,x}^{x,y}$ tends to  
another constant which is too  
complicated to be recorded here.  
  
The last item in this category is $G_{1,x}^{x,x}$. For $0<x<x_1$  
we find  
\begin{eqnarray}  
G_{1,x}^{x,x}=\frac{u}{w^2}\left (\frac{1}{x\hat p^3}-  
\frac{5}{2}\frac{1}{x^2\hat p^2}\right )\quad , \quad p\to 0.  
\end{eqnarray}  
  
Now we list a number of  results for the ``third kind" of  
propagator components with neither of  
their lower indices equal to 1.  
  
Let us start with $G_{x,x}^{x,x}$. For a fixed $x\in (0,x_1)$ 
and $p\to 0$ we find:  
\begin{eqnarray}  
G_{x,x}^{x,x}=\frac{u}{w^2}\left (\frac{1}{x\hat p^3}-  
\frac{7}{2}\frac{1}{x^2\hat p^2}\right )\quad , \quad  
0<x<x_1\quad ,\quad p\to 0.  
\end{eqnarray}  
  
Next we consider $G_{x,x}^{x,y}$, $0<x\le y$. For $0<x<y\le x_1$  
we find a qualitatively new  
type of behaviour:  
\begin{eqnarray} G_{x,x}^{x,y}&=&\frac{2u}{w^2}\left  
(\frac{1}{x\hat p^2}-\frac{7}{2}\frac{1}{x^2\hat p}\right  
)\exp\left [ -\frac{x_1-x}{\hat p}\right ] \quad , \quad  
0<x<y=x_1\quad ,\quad p\to 0,\\  
 G_{x,x}^{x,y}&=&\frac{u}{w^2}\left (\frac{1}{x\hat p^3}-  
\frac{7}{2}\frac{1}{x^2\hat p^2}\right )\exp\left [ -\frac{y-  
x}{\hat p} \right ]\quad , \quad 0<x<y<x_1\quad ,\quad p\to 0.  
\end{eqnarray}  
  
In the long wavelength limit, $p\to 0$, Eqs.~125 and 126 become  
$\delta $-like distributions. Distribution-like  
components were first identified among the spin glass  
propagators by two of us [48], see also [49].  
Independently, Ferrero and Parisi put them to use in their  
recent analysis of the IR divergences in spin glasses [24].  
  
Returning to the list of propagators we find that the component  
$G_{x,y}^{x,y}$, $0<x<y\le x_1$, is equal to the constant in  
Eq.~122 for $0<x<y=x_1$, and  
\begin{eqnarray}  
G_{x,y}^{x,y}&=& 
%\frac{u}{w^2}\left (\frac{1}{x\hat p^3}-  
%\frac{7}{2}\frac{1}{x^2\hat p^2}\right )\exp\left [ -\frac{y-  
%x}{\hat p} \right ] 
-\frac{8u}{w^2}\frac{1}{y^2(y^2-x^2)}\frac{S(x/y)}{S(1)}
\quad , \quad 0<x<y<x_1\quad  
,\quad  p\to 0. 
\end{eqnarray} 
Furthermore, 
\begin{eqnarray} 
 G_{x,y}^{x,x}&=&\frac{u}{w^2}\left (\frac{1}{x\hat p^3}-  
\frac{5}{2}\frac{1}{x^2\hat p^2}\right )\quad , \quad  
0<x<y\le x_1\quad ,\quad p\to 0,\\ 
G^{x,x}_{y,y}&=&\frac{u}{w^2}\left (\frac{2}{y\hat p^2}- 
\frac{7}{y^2\hat p}\right )\exp \left [-2\frac{x_1-y}{\hat 
p}\right ]\quad , \quad y<x=x_1\quad  
,\quad  p\to 0,\\ 
G^{x,x}_{y,y}&=&\frac{u}{w^2}\left (\frac{1}{y\hat p^3}- 
\frac{7}{2}\frac{1}{y^2\hat p^2}\right )\exp \left [-2\frac{x-y}{\hat 
p}\right ]\quad , \quad y<x<x_1\quad  
,\quad  p\to 0. 
\end{eqnarray}

The remaining components depend on three different overlap variables:  
\begin{eqnarray}  
 G_{y,z}^{x,x}&=&\frac{u}{w^2}\left (\frac{1}{x\hat p^3}-  
\frac{1}{2}\frac{1}{x^2\hat p^2}\right )\quad , \quad 0<x<y,z\le  
x_1\quad ,\quad p\to 0,\\[2mm]
G_{x,z}^{x,y}&=&-\frac{4u}{w^2}\frac{1}{x_1(x_1^2-  
x^2)}\frac{S(x/x_1)}{S(1)}\quad , \quad 0<x<y=z=x_1\quad ,\quad  
p\to 0,\\[1mm]
G_{x,z}^{x,y}&=&
-\frac{16u}{w^2}\frac{\hat p}{z^2(z^2-x^2)}\frac{S(x/z)}{S(1)}
%\frac{u}{w^2}\left (\frac{2}{x\hat  
%p^2}-\frac{3}{x^2\hat p}\right )
\exp\left [ -\frac{x_1-z}{\hat p}\right ]  ,\!\!\!\! \quad 0<x<z<y=x_1
,\!\!\! \quad p\to 0,\\[1mm]
G_{x,z}^{x,y}&=&
-\frac{8u}{w^2}\frac{1}{z^2(z^2-x^2)}\frac{S(x/z)}{S(1)}
%\frac{u}{w^2}\left (\frac{1}{x\hat  
%p^3}-\frac{3}{2}\frac{1}{x^2\hat p^2}\right )
\exp\left [ - \frac{y-z}{\hat p}\right ]  , 
\!\!\!\! \quad 0<x<z\leq y<x_1\quad \! ,  
\!\!\!\!\! \quad p\to 0,\\[2mm]
G_{z,z}^{x,y}&=& \frac{u}{w^2}\left (\frac{2}{z\hat  
p^2}-\frac{7}{z^2\hat p}\right )\exp\left [ -2  
\frac{x_1-z}{\hat p}\right ] 
 \quad , \quad z<x=y=x_1\quad ,\quad p\to 0, \\[1mm]
G_{z,z}^{x,y}&=& \frac{u}{w^2}\left (\frac{2}{z\hat  
p^2}-\frac{7}{z^2\hat p}\right )\exp\left [ -  
\frac{x_1+x-2z}{\hat p}\right ]\!\! 
 \quad ,\!\! \quad z<x<y=x_1\quad ,\quad p\to 0, \\[1mm]
G_{z,z}^{x,y}&=& \frac{u}{w^2}\left (\frac{1}{z\hat  
p^3}-\frac{7}{2}\frac{1}{z^2\hat p^2}\right )\exp\left [ -  
\frac{y+x-2z}{\hat p}\right ]\!\!\! 
 \quad ,\!\!\! \quad z<x<y<x_1\quad \!\!,\quad \!\! p\to 0,  
\end{eqnarray}  
and finally $G_{x,z}^{x,y}$ goes to different constants as $p\to  
0$, for $x<y=z=x_1$, $x<y<z=x_1$, and $x<y<z<x_1$, respectively.  
  
The components with one or two upper indices equal to zero are  
special. If the two upper indices  
are different and the smaller is zero, then the propagator  
component  in question vanishes  
indentically. If the two upper indices coincide and vanish, then  
the leading infrared behaviour in the  
diagonal component is given by  
\begin{eqnarray}  
G_{1,1}^{0,0}=\frac{\pi}{4}\frac{u}{w^2}\frac{1}{\hat p^4}\quad , 
\quad p\to 0.  
\end{eqnarray}  
As long as the lower indices $s$, $t$ are larger than zero, the  
off-diagonal components $G_{1,t}^{0,0}$ and $G_{s,t}^{0,0}$  are  
the same as Eq.~138; for $t\to 0$ (or $s$ and/or $t\to 0$) they  
vanish again.

The strong, $p^{-4}$ like singularity in a correlation function
related to Eq.~138 was first noticed by Sompolinsky and
Zippelius [50]. Several results concerning the far infrared
region were published by us [27$-$30], by Goltsev [51],
and by Ferrero and Parisi [24].

The compilation of Eqs.~117$-$138 above gives more details
than any of these papers, nevertheless it  
is far from exhaustive. In a propagator  
component depending on, say, two
% replica indices  
overlap variables
we may have, e.g., that one of them is of the order of $x_1$,  
while the other is much less than even $x_1^2$, thereby defining  
a new mass scale. Depending on the value of the momentum  
relative to these new mass regions we may have further details  
in the IR behaviour. With the exception of Eq.~138 the results 
listed  
in this section have been derived under the assumption that all  
overlap variables are of the same order of magnitude as $x_1$ and 
the momentum is much smaller than $x_1^2$.

\section{The physical meaning of the propagators}  
  
In addition to their role as building blocks  of the  
interacting field theory, the Gaussian propagators have also a  
direct physical meaning which we would like to discuss now.  
This will also allow us to give a physical interpretation to some 
of the results we have found so far.  
  
The propagators are related, as in any field theory, to  
some correlation functions. These correlation functions   
reflect the underlying assumption we have adopted about the  
structure of phase space, namely that it splits into  a large  
number of  equilibrium states with a hierarchical organization.  
  
The simplest correlation function one can define is the overlap  
of the spin-spin correlation function $<s_is_j>$  in valley  
$a$  with the same in valley  $b$:  
\begin{eqnarray}  
C_{ab}({{\bf r}})=\frac{1}{N}\sum_i<s_is_j>_{a}<s_is_j>_{b}=q^2_{ 
ab}+\frac{1}{N}\sum _{\vec {p} }{{\rm e}}^{-{{\rm i}}\vec {p}  
\vec {r}}C_{ ab }(\vec {p} )  
\end{eqnarray}  
where the distance   ${{\bf r}}={{\bf r}}_i-{{\bf  
r}}_j$  between the sites is kept fixed  as we sum over  
the lattice points $i$, and $q_{ab}$  
is the overlap between states $a$, $b$, Eq.~31.  
  
In principle,  $ C_{ab}({{\bf r}})$ as defined in  
Eq.~139 could depend on  the concrete realization of the random  
couplings $J_{ij}$  and also on the two states $a$ and $b$. 
Following the considerations of Parisi [38]  
and also those of M\'ezard and Virasoro [52] one is led to the  
conclusion, however, that $C_{ab}({{\bf r}})$ is  
self-averaging (i.e.\ independent of the sample $J_{ij}$ for a  
large system) and depends on  $a$ and $b$  
only through the overlap $q_{ab}$. Moreover, the  
Fourier transform  
$$C_{ab}(\vec {p} )=\frac{1}{N}\sum _{ij}\exp[{{\rm  
i}}({{\bf r}}_i-{{\bf r}}_j)\vec {p} 
]<s_is_j>_{a}<s_is_j>_{b}-Nq^2_{ab}\delta _{\vec {p},0}^{Kr}$$  
can be calculated in the Gaussian approximation via the replica  
formalism and turns out to be nothing but the diagonal  
component of the propagator:  
\begin{eqnarray}  
C_{ab}(\vec {p} )=G_{1,1}^{x,x}(\vec {p} )  
\end{eqnarray}  
where the relationship between $a$, $b$ and $x$ is  
given  
by the inverse of Parisi's order paramater function: 
$x=x(q_{ab}$). The derivation of Eq.~140  
follows the same steps that led Parisi to the identification of  
$\displaystyle \frac{{{\rm d}}x}{{{\rm d}}q}$ as the  
probability  
distribution of overlaps [38], and can  
therefore be omitted here. The  essential ingredient is the  
assumption about the states having the property of clustering.  
  
Correlation functions involving three or four states can be  
treated similarly:  
\begin{eqnarray}  
C_{abc}({{\bf r}})=\frac{1}{N}\sum _i<s_is_j>_{a} 
<s_i>_{b}<s_j>_{c} = q_{ab} q_{ac}+\frac{1}{N}\sum _{\vec {p} 
}{{\rm e}}^{- {{\rm i}}\vec {p} \vec {r} }C_{abc}(\vec {p} )  
\end{eqnarray}  
and  
\begin{eqnarray}  
C_{abcd}({{\bf r}})=\frac{1}{N}\sum _i<s_i>_a<s_i>_b 
<s_j>_c<s_j>_d=q_{ab}q_{cd}+\frac{1}{N}\sum _{\vec {p}  
}{{\rm e}}^{-{{\rm i}}\vec {p} \vec {r} }C_{abcd}(\vec {p} )  
\end{eqnarray}  
are also self-averaging and depend on the states $abc$,
resp.~$abcd$ only through the overlaps.
The Fourier transforms are 
given by  
the off-diagonal components of the propagators as  
\begin{eqnarray}  
C_{abc}(\vec {p})=G_{1,z}^{x,y}(\vec {p} )  
\end{eqnarray}  
with $x=x(q_{ab})$, $y=x(q_{ac})$, 
$z=\max\{x(q_{ba}),x(q_{bc})\}$, and  
\begin{eqnarray}  
C_{abcd}(\vec {p} )=G_{z_1,z_2}^{x,y}(\vec {p} )  
\end{eqnarray}  
with $x=x(q_{ab})$, $y=x(q_{cd})$, $z_1=\max \{x(q_{ac}), 
x(q_{ad})\}$, $z_2=\max \{x(q_{bc}), x(q_{bd})\}$. 
  
Eqs.~140, 143 and 144, first written up in [30], give a  
direct physical meaning to the various propagator components.  
We see now that the complicated behaviour of the propagators  
reflects the structure of correlation overlaps inside a state  
and also between different states.  
  
In particular, we can see that the propagator components with  
all overlaps above $x_1=x(q_{{\rm max}})$ correspond indeed to  
correlation  
functions inside a single state, as anticipated in Sec.~6.
In the near  
infrared region we  found for these propagators the simple  
forms in Eq.~58, and the transverse and longitudinal combinations 
given  
in  Eq.~59. These clearly suggest the idea of a massless  
phase with a transverse correlation function falling off like  
$1/r^{d-2}$ in real space and an exponentially decaying  
longitudinal correlation function with a characteristic length  
\begin{eqnarray}  
\xi=\frac{1}{\sqrt{2wq_1}}\approx \frac{1}{\sqrt {2\tau} }\quad .
\end{eqnarray}  
This reveals the meaning of the large mass: it is the inverse   
coherence length of the longitudinal fluctuations in a single  
state.  
  
In order for this interpretation to be consistent, the  
single-state propagators should not develop stronger infrared  
singularities in the far infrared region than the behaviour we  
have just seen in the near infrared. Recalling Eqs.~117, 118
and 119, valid for momenta far below the upper edge of the  
small mass band, we see that this is precisely what happens: the  
leading terms $3/p^2$, $3/2p^2$ and $1/p^2$  
one finds for the single state propagators in the far  
infrared coincide with  the $p\to 0$  
limit of the formulae (58), derived in the near  
infrared. It seems therefore that the long-distance behaviour  
of the single-state propagators is relatively mild: although  
the  
phase is massless, the infrared behaviour does not appear to be  
more violent than in the Heisenberg model, for example. In  
addition, as shown in [42], the result $1/p^2$  for the  
transverse  
single-state propagator is, in fact, exact within the Gaussian  
approximation; it holds not only in the truncated model or  
near $T_c$, but throughout the ordered phase.

For a complete characterization of the correlations in the  
system we have to study also the interstate overlaps of  
correlation functions, with some or all replica overlaps going  
below $x_1$. The behaviour one finds in this case depends  
radically  
on the momentum range considered. For momenta in the near  
infrared the correlation functions display a reasonable degree  
of complexity: $p$ scales with the large mass, and the overall  
scaling power is $1/p^2$, as shown by Eqs.~60$-$69. This  
means  
that the qualitative features of the correlation overlaps  
between  
two states with a given overlap $q<q_{{\rm max}}$ are similar  
to  
those inside a single state, except on the extremely long  
wavelength scales. The characteristic length beyond which  
qualitative changes occur is $\xi '\sim 1/\tau $, 
corresponding  
to the upper edge of the small mass band. This gives a meaning to 
the small mass scale.  
  
As we see from the results in the previous section, for 
wavenumbers  
in the range of the small mass or below, the correlation overlaps 
show very strong infrared divergences, going up to 
${1/p^4}$. This violent IR  behaviour poses a formidable 
challenge to spin glass theory.

\section{The first-loop corrections above 8 dimensions} 
  
In this section we apply the previous results  for the 
calculation of the first short-range corrections  
to the equation of state and to the mass spectrum. The treatment 
follows our previous works [15,16] closely.  
For the time being we assume that the dimensionality is 
sufficiently high for the ordinary loop-expansion to work: as we 
shall see shortly, this means $d>8$.  
  
Our starting point is Eq.~18 where we have to remember that 
$q_{\alpha \beta }$ is, in principle, the exact order  
parameter, but $\tau $ is the reduced temperature measured 
relative to the mean field transition  
temperature. Let $\tau _c$ be the critical value of $\tau $ where 
the exact $q_{\alpha \beta }$ vanishes. Now we can write up  
Eq.~18 at $\tau _c$ with the help of Eqs.~40 and 41 which are valid 
for $q_{\alpha \beta }\to 0$, and subtract the resulting equation 
from (18). We get:  
\begin{eqnarray} 
&&2tq_{\alpha \beta }+w(q^2)_{\alpha \beta 
}+\frac{2u}{3}q_{\alpha 
\beta }^3+\\ 
&&+\frac{1}{z}\frac{1}{(2\pi )^d}\displaystyle\int _{|\vec 
p|<1}^{} 
\!\!\!\! 
{{\rm d}}^d p\left [ w \sum_{\gamma \ne \alpha ,\beta }\left 
(G_{\alpha 
\gamma  ,\gamma  \beta }(\vec {p})-\frac{wq_{\alpha \beta 
}}{p^4}\right )+2uq_{\alpha \beta }\left (G_{\alpha \beta ,\alpha 
\beta }(\vec p)-\frac{1}{p^2}\right )\right ]=0 \quad \!\!\! 
, 
\nonumber \end{eqnarray} 
where $t=\tau -\tau _c$ is the ``true" reduced temperature. 
The critical value of $\tau $ works out to be  
\begin{eqnarray} 
\tau _c=\frac{1}{z}(w^2I_4-uI_2)\quad , 
\end{eqnarray} 
where  
\begin{eqnarray} 
I_l=\int _{|\vec p| <1} \frac{{{\rm d}}^dp}{(2\pi 
)^d}\frac{1}{p^l}\quad , \quad l=2,4,6,\ldots\quad . 
\end{eqnarray} 
(For $l\ge d$ an IR cutoff  at $p>\sqrt{2t}$ is understood in the 
integrals $I_l$.)  
  
Since $\tau _c$ is of ${{\cal O}}(1/z)$ we can replace $\tau $ by 
$t$ in the propagators appearing in the loop terms. These  
propagators satisfy the set of equations (36)$-$(39), with $\tau $ 
replaced everywhere by $t$. In the  
dimensionality range considered here the dominant contributions 
to the loop integrals come from  
the neighbourhood of the upper cutoff. In this momentum range the 
``large-$p$ expansion" results of Eqs.~42$-$44 are valid. As the mean 
field part of the equation of state goes up to ${{\cal O}}(q^3)$, we have 
to stop at the  same order also in the loop terms. This means we 
have to go to ${{\cal O}}(q^3)$ in $G_{\alpha \gamma ,\beta \gamma }$ in 
Eq.~43, but have to stop at ${{\cal O}}(q^2)$ in $G_{\alpha \beta ,\alpha 
\beta }$, Eq.~42, which is already multiplied by $q_{\alpha 
\beta }$. Substituting these approximate propagators  
into Eq.~146, performing the summation over the replica index 
$\gamma 
$ and collecting the coefficients of  $q_{\alpha \beta }$, 
$(q^2)_{\alpha \beta } $, and  $q^3_{\alpha \beta }$,
we end up with  an equation which is of the same form as the mean 
field equation of state with $t$, $w$ and $u$ replaced by some 
new, renormalized coupling constants $\widetilde t$, $\widetilde 
w$ 
and $\widetilde u$. The mapping works out  
as follows: 
\begin{eqnarray} 
\begin{array}{rcl} 
 \widetilde t &=& t \left (1 - \displaystyle 
\frac{1}{z}[4w^2I_6+12w^2tI_8-2uI_4-4utI_6]\right ) \\  
 \widetilde w &=& w \left (1 - \displaystyle 
\frac{1}{z}[2w^2I_6+12w^2tI_8]\right ) \\  
 \widetilde u &=& u \left (1 - 
 \displaystyle\frac{1}{z}[12w^2I_6- 
\frac{12w^4}{u}I_8-6uI_4]\right )\quad . 
\end{array} 
\end{eqnarray}

In addition to the above, some new type of couplings are also 
generated by the loop terms. These  
would be corrections to those quartic couplings which we 
discarded already at the very beginning.  
Since they enter only subleading terms in all quantities which we 
are interested in, we can safely  
ignore them.  
  
With this we have mapped back our loop-corrected theory onto MFT. 
The order parameter will   
be of the same form as Parisi's order parameter function given in 
Eqs.~27 and 28, with $\tau $ replaced by $\widetilde t$  
and $w$, $u$ replaced by $\widetilde w$, $\widetilde u$, 
respectively. A more detailed discussion of the effects of these 
replacements can be found in [15,16], here we  merely note that the 
slope $\widetilde w/2\widetilde u$ of $q(x)$ decreases, while the 
breakpoint $x_1$  
increases with decreasing dimensionality. This means that replica 
symmetry breaking effects are  
enhanced by the loop corrections, in complete agreement with the 
conclusions drawn by Georges,  
M\'ezard and Yedidia from the $1/d$ expansion [14].
 
We can arrive at the same conclusion by considering the 
excitation spectrum of the system. In order to be able to 
calculate the renormalized masses we need first the corrections 
to the self-energy matrix. At the first loop level these are 
typically bilinear combinations of the various propagator 
components, integrated over the loop momentum. For our present 
purposes we may disregard ``wave function renormalization  
effects" and evaluate the self-energies for zero external 
momentum. For $d>8$ the propagators entering the calculation can 
again be replaced by their ``large-$p$" approximations from 
Eqs.~42$-$44 where we can now stop at ${{\cal O}}(q^2)$. Performing all the 
necessary replica summations and momentum integrations we find at 
the end that the self-energies to ${{\cal O}}(1/z)$ are precisely of the 
same form as at the zero loop level, Eq.~85, with the 
replacement of $\tau $, $w$ and $u$ by the renormalized coupling 
$\widetilde t$, $\widetilde w $ and $\widetilde u$, respectively. 
This shows the full consistency of our scheme and guarantees that 
important qualitative features of the spectrum (e.g.\ gaplessness) 
are not violated. The renormalized masses themselves can then be 
obtained from their mean-field counterparts, with the replacement 
of the bare couplings by the tilded ones. Considering Eq.~109 
we can then see that the centre of the large mass band will 
change very little but, due to the renormalization of the $u$ 
coupling, the bandwidth will grow considerably as the dimension 
$d$ decreases. The same is true for the upper edge of the small 
mass band. These are further signals of growing RSB effects 
(without RSB the spectrum would consist of two points).  
 
We can sum up these findings in the following way: Parisi's RSB 
field theory has proved to be perturbatively stable above $d=8$ 
dimensions. The loop corrections do not change the qualitative 
features of the theory, they just shift the constants by a small 
amount, and, at least to the low order regarded here, they even 
enhance RSB effects. 
 
As we approach $d=8$ from above, however, the mapping set up in 
Eq.~149 becomes singular: the integral $I_8$ (cut off at the lower end at 
$p=\sqrt{2t}$) blows up. This causes the most serious problem in 
$\widetilde u$ where $I_8$ appears alone, without any additional 
$t$ factor multiplying it.    As it stands, $\widetilde u$ simply 
does not make sense below $d=8$: for $t\to 0$ it diverges 
like $t^{(d-8)/2}$. The next section is devoted to the resolution 
of this paradox.

\section{Between 6 and 8 dimensions}
 
The formal reason for the mapping (149) becoming singular is easy
to understand: whereas for $d>8$  
the loop integral $I_8$ is dominated by the contribution of
momenta around the UV cutoff, for $d<8$ the  
main contribution comes from the lower end where the momentum is
of the same order of  
magnitude as the large mass. Then the large-$p$ expansion is not
a good representation of the  
propagators any more. In particular, as the $p^{-8}$ singularity
comes from the expansion of $G_{\alpha \gamma ,\gamma \beta}$, we
have  to treat this component more carefully. (The other
propagator, $G_{\alpha \beta ,\alpha \beta }$, in the $u$-loop
term in (146) does  not cause problems for $d>6$, so we can
continue to use the large-$p$ expansion result Eq.~42 for it.) 

Working out the replica summation in Eq.~146 we get in the
continuum limit: 
\begin{eqnarray}
\sum_{\gamma \ne \alpha,\beta }G_{\alpha \gamma ,\gamma \beta
}= -\int _0^x{{\rm d}}yG_{x,1}^{y,y}-xG_{x,1}^{x,x}-
2\int_x^1{{\rm d}}yG_{x,1}^{x,y}\quad ,\quad \alpha \cap \beta
=x\quad .
\end{eqnarray}
 
Considering that $x\le x_1\sim t \ll 1$, it is clear that the
largest term in Eq.~150 comes from the upper end of the last  
integral, so we can again use the ``innermost block approximation"
here: 
\begin{eqnarray}
\sum _{\gamma \ne \alpha ,\beta }G_{\alpha \gamma ,\gamma \beta }
\approx -2G_{x,1}^{x,x_1}\quad .
\end{eqnarray}
 
Now one can check that for $d>6$ the contribution of the far
infrared region to the integral of Eq.~151  
over the momentum is still negligible, so we can use Eq.~62, valid
in the near infrared, for the  
propagator in Eq.~151. Then substituting Eqs.~62 (taken at $y=x_1$)
and 42 into Eq.~146 and collecting the  coefficients of $q(x)$
and $q^3(x)$, respectively, we arrive finally at a new set of
renormalized coupling  
constants which are, in fact, very similar to those in Eq.~149. The
only difference is that  $\widetilde u$ is now  
given by 
\begin{eqnarray}
\widetilde u= u+12\frac{w^4}{z}\int \frac{{{\rm d}}^dp}{(2\pi
)^d}\frac{1}{p^4(p^2+2t)^2}\quad ,
\end{eqnarray}
where $2wq_1$ in the denominator of the integrand has been
replaced by its zeroth order value $2t$. In the  
dimension range $6<d<8$ regarded here the integral in Eq.~152 is
well-defined both for small and for large  
momenta, so to leading order in $t$ we can send the upper cutoff
to infinity. The effective coupling  
$\widetilde u$ then becomes 
\begin{eqnarray}
\widetilde u=u+{{\rm const}}\cdot \frac{t^{(d-8)/2}}{z}\quad .
\end{eqnarray}
for fixed $t$ and $z$ very large (i.e.\ for temperatures not very
close to $T_c$ and for very long range  
forces) the shift in the quartic coupling is small, and we are
back to the previous situation: the loop-corrections do not
significantly alter the mean field results. In the opposite
limit, however, where  
$z$, albeit large, is kept fixed and the temperature is allowed
to go arbitrarily close to $T_c$, the  
bare $u$ becomes negligible compared with what was supposed to be
a small correction. It  is clear  
that under these circumstances the ordinary loop-expansion breaks
down. 
 
Standard power counting arguments tell us that the upper critical
dimension of the model defined in  
Eq.~2 is $d_c=6$. It would therefore be perfectly normal to find an
infrared breakdown of the perturbation  
expansion in six dimensions. Why we should have IR problems
already in $d=8$, however, demands
explanation.

To understand the origin of the problem, we have to realize that
the term that blows up in $8$  
dimension is a one-loop contribution to the $4$-point function at
zero external momenta. It is evident  
that this quantity (the ``box graph") should be singular at the
critical point in $d=8$. 
 
It should be emphasized that this effect is not a spin glass
peculiarity. Many-point functions at  
exceptional momenta (where power counting arguments fail) can
blow up at $T_c$ in high dimensions in  
any theory. For example, the one-loop correction to the sixth
derivative of the thermodynamic  
potential with respect to the magnetization in an ordinary
$\varphi ^4$ theory (the triangle graph) will be  
singular at $T_c$ in $d=6$, i.e.\ {\it above} the upper critical
dimension of that model. This six-point function  
is related to a higher order nonlinear susceptibility, which is
normally of little interest, so the  
problem is rarely mentioned. The triangle graph (and all the
higher polygons that have similar IR  
singularities in higher and higher even dimensions) enter,
however, as insertions also in some of the  
high order loop-corrections to physical quantities such as the
two-point function (i.e.\ the inverse  
linear susceptibility) which are regularly discussed and claimed
to be free of IR problems above the  
upper critical dimension. One may wonder whether these
IR-singular many-point insertions do not  
make also the 2-point function singular. They do not: the
singularity in the many-point functions  
occurs only at exceptional (zero external) momenta, so when
these graphs appear as insertions in  
higher order diagrams and finite momenta flow through them their
singularity will be suppressed by  
the loop integrals. Therefore, the isolated IR singularities
appearing in the many-point functions do  
not proliferate in higher orders and do not destroy  the
loop-expansion for the quantities which  
one is usually interested in. 
 
The situation in spin glass field theory is quite similar to the
above, except that here the many-point  
functions do appear
directly in the coefficients of the Taylor 
expansion of the order parameter in the variable $x$. In
particular, the 4-point function appears in the $x^3$
term in the equation of state, hence in the slope of $q(x)$,
so if we want to calculate  
$q(x)$ to linear order we have to learn how to handle this
difficulty. 
 
Since above 6 dimensions the singularity will not proliferate in
higher orders, it is evident what we  
have to do. We have to absorb this isolated singular term into
the mean field part of our equation  
and treat the rest as perturbation. This way we can save the
loop-expansion between 6 and 8  
dimensions in a reorganized form. 
 
The starting point in this new expansion is an effective mean
field theory which is exactly of the  
same structure as what we have discussed so far but with the bare
coupling  $u$ replaced by $\widetilde u$  
everywhere. Since the remaining loop-corrections do not
qualitatively modify the results, in the  
following we restrict ourselves to the discussion of this
effective MFT. Moreover, we will be  
interested only in the fluctuation dominated regime where 
\begin{eqnarray}
\frac{t^{(d-8)/2}}{z}\gg 1\quad ,
\end{eqnarray}
so the bare $u$ can even be completely omitted. 
 
The nontrivial renormalization of the four-point coupling has a
profound consequence for the  
structure of the theory. To fully appreciate its significance we
have to go back for a moment to the  
original MFT. As first stressed by Fisher and Sompolinsky [53],
scaling is badly violated in Parisi's  
MFT: not only the so called hyperscaling laws break down, but
also those that do not explicitly  
contain the dimension $d$, such as e.g.\ $\beta \delta =1-\alpha
/2+\gamma /2$. As a matter of fact, it is not even possible to
unambiguously assign a critical exponent  to some physical
quantities. For example, the maximum of the order  
parameter function scales as $q_1\sim t$, but Sompolinsky's
susceptibility anomaly [54] $\Delta=
\frac{1}{T}\left (q_1-\int_0^1{{\rm
d}}xq(x)\right )$ which is a  particularly useful measure of
ordering in the presence of an external magnetic field goes as
$\Delta \sim t^2$. The  existence of two ``mass scales", i.e.\ two
characteristic lengths, one diverging as $t^{-1/2}$  the other as
$t^{-1}$,  
is another example of this ambiguity. Now the most important
effect of the replacement $u\to \widetilde u$ below 8 
dimensions is that upon approaching the upper critical dimension
$d=6$ from above scaling  gets  gradually restored [53], [15,16]. 
 
Indeed, consider e.g.\ the slope $c=w/2\widetilde u$ of $q(x)$.
With $\widetilde u\sim t^{d/2-4}/z$, valid in the fluctuation 
dominated regime,  the slope becomes $c\sim zt^{4-d/2}$. For $d
\to 6^+$ $c$ is therefore $\sim t$. On the other hand, the
plateau of $q(x)$ is independent  of $\widetilde u$, so it
remains $q_1 \sim t$. As for the  
breakpoint $x_1=2 \widetilde u t/w^2$, it becomes $\sim t^{d/2-
3}/z$. Note that for $d\to 6^+$ the breakpoint becomes {\it
independent} of the  
temperature (but remains small, of the order of $1/z$). What all
this amounts to is that upon  
approaching $d=6$ from above the order parameter function becomes
of the form 
\begin{eqnarray}
q(x)=t^{\beta }f(x)\quad ,
\end{eqnarray}
where $\beta =1$, and $f(x)$ is independent of temperature. The
form (155) is necessary for all the meaningful  
combinations one can form from $q(x)$ to scale with the {\it
unique} exponent $\beta $. 
 
Similar remarks apply for the mass scales. The center of the band
of the large eigenvalues remains  
$2t$, but as shown by Eq.~109 the band width $4t^2\widetilde
u/3w^2$ varies with  $d$ as $\sim t^{d/2-2}/z$ and becomes $\sim
t/z$ as $d\to 6^+$. The same is  true for the upper edge
$2\widetilde ut^2/w^2$ of the band of small eigenvalues. This
means that instead of two mass  
scales having different temperature exponents, we have a single
exponent $\nu =1/2$ for $t\to 6^+$. Note however  
that the separation of scales still remains in the form of a
numerical difference: the small  
eigenvalues and the  band width of the large eigenvalues is down
by a factor $1/z$ compared with  
the center of the large eigenvalues. 
 
To make contact with the work of Green, Moore and Bray [55] (who
were, in fact, the first to note  
the role of the renormalized four-point coupling in a particular
instance), we consider now the effect  
of  an external field $h$ on the order parameter $q(x)$. Although
we have avoided the problem of the  
field in this paper, it is easy to show that the replacement $u
\to \widetilde u$  will work also in the presence of the  
field. Therefore the well-known results for the field dependence
of $q(x)$ can be readily taken over  
from ordinary MFT, and one can see, in particular, that the
%de Almeida-Thouless
AT line [31] will be  
given by $h^2_{AT}=4\widetilde ut^3/3w^3$. The standard mean
field result $h^2_{AT}\sim t^3$  will then be replaced by
$h^2_{AT}\sim t^{d/2-1}/z$ for $6<d<8$, becoming $h^2_{AT}\sim
t^2/z$ as $d\to 6^+$. This coincides with the result found by
Green et al.\ [55] which they obtained from the zero of the  
replicon self-energy in a one-loop calculation in the disordered
phase. The agreement between  
these two independent calculations is a testimony for the
consistency of the effective MFT. Green et  
al.\ also suggested that the shifted AT exponent might be exact.
This must  indeed be true for all the  
exponents predicted by the effective MFT, because, as argued
above, the singularity found in the  
four point function will not proliferate and the higher loop
corrections will not be more singular then  
the one-loop contribution worked out here. 
 
When we were talking about the limit $d\to 6^+$ we always meant
to remain slightly above $d=6$. In exactly six  
dimensions the other corrections, disregarded in the effective
MFT also become singular and  
produce the usual logarithms appearing at the upper critical
dimension. The little we know about  
the range $d<6$ is the subject of the next section. 
 
\section{The first step below 6 dimensions}
 
As we cross $d=6$ infrared divergences start to plague the whole
perturbation expansion. Now they 
do proliferate: every order will be more singular then the
previous one. In this situation one 
normally turns to renormalization group (RG) methods to
reorganize the perturbation expansion. The 
expansion in the distance  $\varepsilon$ from the upper critical
dimension proved particularly successful in the 
context of ordinary critical phenomena. The application of field
theoretic RG methods to spin  glasses was pioneered by
Harris, Lubensky and Chen 
[32]. Building on [33],
Green calculated spin glass critical exponents
to ${{\cal O}}(\varepsilon ^3)$ in [34].
The structure of the RG in spin 
glasses is poorly understood, however. No systematic RG analysis
has ever been made in the 
condensed phase, and the negative result of Bray and Roberts [56]
who failed to locate a stable fixed 
point in a field below $d=6$ remains an enigma to resolve. 
 
Neither do we have a magic key to the riddle. What we are able to
do here is to take just the very 
first step toward an $\varepsilon $-expansion in the ordered
phase of spin glasses, without the solid basis of an 
RG analysis. The findings in the previous section provide some
important clues. 
 
The first of these is that the bare quartic coupling $u$ can be 
dropped altogether $-$ the cubic theory 
builds up its own quartic coupling at the first loop level.
This was already observed by Bray and Moore [57]
who, working around a replica symmetric starting point,
run into an AT-like instability, however.
Secondly, with the restoration of scaling 
the breakpoint $x_1$ of the order parameter becomes
small, of  
${{\cal O}}(1/z)$. If there exists an $\varepsilon $-expansion at
all, this should translate into $x_1={{\cal O}}(\varepsilon )$
below $d=6$. But if $x_1$ is small, so is the whole region where
RSB occurs, hence to leading order we can apply the 
innermost block approximation in all replica summation. Thirdly,
the upper edge of the small mass 
band must also become ${{\cal O}}(\varepsilon )$ , so to leading
order we can  always use the near infrared propagators. 
 
With all these simplifications the calculation of some of the
critical exponents to ${{\cal O}}(\varepsilon )$ becomes quite 
feasible, for the exponent $\beta $ which we present here as an
illustration, it becomes almost trivial. 
 
In order to obtain $\beta $, it is sufficient to calculate $q_1$.
Let us regard the equation of state as a 
polynomial in $x$. Then a little reflection shows that the
maximum of $q(x)$ is determined by the linear 
terms in $x$. (The slope at $x=0$ would be determined by the
$x^3$ terms.) 
 
Where do we get linear terms in Eq.~146? The first term is such. In
the second term we have  
\begin{eqnarray}
(q^2)_{\alpha \beta }=-\int _0^x{{\rm d}}yq^2(y)-xq^2(x)-
2q(x)\int_x^1{{\rm d}}yq(y)\quad ,\quad \alpha \cap \beta =x\quad
,\end{eqnarray}
which provides a linear term coming from the last integral 
\begin{eqnarray}
(q^2)_{\alpha \beta }\to-(2-x_1)q_1q(x)+{{\cal O}}(x^3)\quad .
\end{eqnarray}
With the bare $u$ omitted the only term we are left with is the
$w$-loop. Using the innermost block 
approximation (151) for the replica summation, and selecting the
linear term from the near infrared 
approximation (62) for the propagator appearing here we have 
\begin{eqnarray}
\sum_{\gamma \ne \alpha ,\beta }\left (G_{\alpha \gamma ,\gamma
\beta }-\frac{wq_{\alpha \beta }}{p^4}\right )=-
\frac{4wtq(x)}{p^2(p^2+2t)^2}+{{\cal O}}(x^3)\quad ,
\end{eqnarray}
where, in view of this loop term being of ${{\cal O}}(\varepsilon
)$, we used the zeroth order result $wq_1=t$.
 
Collecting Eqs.~156$-$158 and evaluating the integral of Eq.~158 in
$d=6$ for $t\to 0$, we find that the equation 
of state to first order in $x$ is: 
\begin{eqnarray}
\left ( 2t-2wq_1+wx_1q_1-\frac{4w^2K_6}{z}\ln \sqrt{2t}\right
)q(x)+{{\cal O}}(x^3)=0\quad ,
\end{eqnarray}
where $K_6=\frac{1}{64\pi^3}$. 
 
Writing $wq_1=At^{\beta }=(1+A^{(1)}+\cdots )t(1+\beta ^{(1)}\ln
t +\cdots )$ where $A^{(1)}$ and $\beta ^{(1)}$ are the ${{\cal
O}}(\varepsilon )$ corrections to the amplitude $A$ and to the
critical exponent $\beta $, respectively, we finally obtain 
\begin{eqnarray}
\beta ^{(1)}=\frac{w^2K_6}{z}\quad .
\end{eqnarray}

From [33,34] we know that at the fixed point of the cubic  coupling
the combination in Eq.~160 is $\varepsilon /2$. Thus 
we get 
$$\beta =1+\frac{\varepsilon }{2}+{{\cal O}}(\varepsilon ^2)$$
which is in accord with the exponents calculated at $T_c$ [34].

If the interpretation of the IR logarithm coming from the
loop-integral as a correction to $\beta $ is correct 
then at the next order the coefficient of $\ln ^2t$ must be half
of the coefficient of the first order log. We have evaluated the
two-loop correction to the equation of state to $\ln ^2 t$ order
and found that  the above condition of exponentiation is
fulfilled, indeed, provided (160) is chosen $\varepsilon /2$, its
fixed point  value. 
 
Although this provides an important check on our
first order calculation and raises the hope that the $\varepsilon
$-expansion can be consistently carried through also below $T_c$,
we have to point out that at the second order we find not only
the $\ln ^2 t$, $\ln t$ terms necessary for scaling, but also
infrared divergences like $p^{-6}$.
%They are the consequences of
%the small masses and, as a matter of fact, were to 
%be expected. Similar IR divergences appear also in the ${\rm
%O}(n)$ model where they can be systematically removed due to the
%rotation symmetry of the model. We have not yet been able to find
%similar way to get rid of the IR problem in spin glass field
%theory. 
In higher orders we will clearly find even stronger infrared powers, due 
to the presence of zero modes and soft modes in the system. Though we 
do notice some cancellations, a systematic method (similar to the 
exploitation of rotational symmetry through the use of Ward identities in 
the ${\rm O}(n)$ model,
or the use of the small masses as an infrared cutoff) is 
still to be found in spin glass theory.
Work is in progress in that direction.

\section{Summary}

Let us briefly recapitulate the main points in the paper.
Having set up the field theoretic formalism, we devoted a long 
discussion to the free, quadratic fluctuations about an equilibrium 
solution with Parisi's ultrametric symmetry. We showed that for momenta 
higher than any characteristic mass, i.e.~for wavelengths shorter than 
any characteristic distance in the system, the propagators describing 
these fluctuations can be obtained even without having to specify the 
concrete symmetry breaking pattern.

In the long but not extremely long wavelength limit, that we called the 
near infrared, we obtained simple approximate forms (rational fractions 
with poles at zero and at the large mass) for the propagators. We also 
found that in this region the propagators show a simple and explicit 
functional dependence on the order parameter: they are polynomials, going 
up to the quartic order, in $q(x)$.
 
We also displayed the group theoretic and algebraic techniques one 
needs to block- diagonalize any ultrametric matrix. The additional 
simplification occuring in the kernel of these block- diagonal forms in 
the case of the Gaussian propagators near $T_c$
then allowed us to obtain exact, closed 
expressions from which the far infrared behaviour could be determined.
The dependence of the propagators on $q(x)$ is much more complicated in 
this region than in the near infrared, in particular, the high IR powers we 
found for extremely small wavenumbers are intimately related to $q(x)$
being a 
linear function for small $x$.

The physical meaning of these results could be understood on the basis 
of the relationships we established between the propagators and some intra- 
and intervalley overlaps of spin-spin correlation functions. In 
particular, we learned that transverse fluctuations are long-ranged also 
inside a single phase, while the longitudinal fluctuations have a finite 
coherence length, given by the inverse large mass. Overlaps of 
correlations between different phase space valleys were found to start to 
qualitatively deviate from those inside a single phase around a 
distance given by the inverse small mass. Most remarkably, inside a 
single phase the extreme long wavelength behaviour was found to be a smooth 
continuation of the behaviour on the intermediate, near infrared scales, 
whereas the overlap between correlation functions in two distant valleys 
shows a markedly different, more singular behaviour in the far infrared 
than in the near infrared.

Having understood the structure of correlations in the various regimes 
we proceeded to apply our propagators in the calculation of the first 
loop corrections. We showed that for $d>8$ the theory maps back onto MFT 
with small, numerical shifts in the coupling constants, which demonstrated 
the stability of Parisi's MFT against short range corrections, at least to 
the low order investigated here.

For $6<d<8$ we had to reorganize the loop expansion in order to incorporate 
a divergent correction to the quartic coupling. This led us to an 
effective MFT with exactly calculable but dimension dependent critical 
exponents, and enabled us to follow the development of the theory with 
decreasing dimensionality towards a form reached at $d=6^+$, with scaling 
restored but a nontrivial replica symmetry breaking pattern preserved.

Entering the range $d<6$ we noted that both the breakpoint $x_1$ and the 
small masses become of the order of $\varepsilon$ . This simplified the 
calculation tremendously, and allowed us to readily identify the term 
that yields the ${\cal O}(\varepsilon )$ correction to the critical
exponent of the 
order parameter, and even to check exponentiation at the next order.

We concluded the paper by pointing out the challenge posed by the 
appearence of infrared singularities in the expansion and by hinting at 
the methods by which this challenge may be met in the future.         

\section*{Acknowledgments}
During the course of the long-term research project reviewed in this 
paper we benefitted from interaction with a large number of colleagues, 
of whom special thanks are due to A.~Bray, E.~Brezin, B.~Derrida, S.~Franz, 
Z.~Horv\'ath, C.~Itzykson, M.~M\'ezard, M.~A.~Moore, H.~Orland,
G.~Parisi, N.~Sourlas,
A.~P.~Young, J.~Zinn-Justin and J.-B.~Zuber.
This work has been supported by the French
Ministry of Foreign Affairs, contract NGE3T0/023,
by the Hungarian National Science Fund OTKA, Grant
No.~T~019422, and by the European Research Network on
``Statistical Physics of 
Collective Behaviour in Disordered Systems and Information Processing",
% PECO .... .
contract CHRX-CT92-0063.

\end{document}